\documentclass[12pt]{article}

\usepackage{amsmath}
\usepackage{amsthm}
\usepackage{amsfonts}          
\usepackage{amssymb}           
\usepackage[mathscr]{eucal}
\usepackage{graphicx}
\usepackage{color}
\usepackage{bbold}

\usepackage[isu,bf]{caption2}
\newlength{\xtrawidth}
\newlength{\xtraheight}
\usepackage[all]{xy}           

\def\texorpdfstring#1#2{#1}

\def\clap#1{\hbox to 0pt{\hss#1\hss}}

\def\mathclap{\mathpalette\mathclapinternal}

\def\mathclapinternal#1#2{%
\clap{$\mathsurround=0pt#1{#2}$}}	

\newcommand{\eqdef}{%
  \mathrel{\lower.1mm
    \hbox{$\stackrel{\lower.424ex\hbox{\scriptsize def}}{=}$}}
}
\newcommand{\Q}{\ensuremath{{\mathbb{Q}}}}
\newcommand{\R}{\ensuremath{{\mathbb{R}}}}
\newcommand{\C}{\ensuremath{{\mathbb{C}}}}

\newcommand{\Z}{\mathbb{Z}}
\newcommand{\CP}{\ensuremath{\mathop{\null {\mathbb{P}}}}\nolimits}

\newcommand{\CY}{Calabi-Yau}
\newcommand{\CYm}{\CY{} manifold}

\newcommand{\MW}{Mordell-Weil}
\newcommand{\MWgrp}{\MW{} group}

\newcommand{\even}{\ensuremath{\mathrm{ev}}}

\newcommand{\Ncal}{\mathcal{N}}

\newcommand{\ptset}{\ensuremath{\{\text{pt.}\}}}

\DeclareMathOperator{\diff}{d}

\DeclareMathOperator{\Pic}{Pic}

\DeclareMathOperator{\tr}{tr}

\DeclareMathOperator{\rank}{rank}

\DeclareMathOperator{\End}{End}

\DeclareMathOperator{\Sym}{Sym}
\DeclareMathOperator{\idx}{Index}

\DeclareMathOperator{\Ext}{Ext}

\DeclareMathOperator{\HOM}{\underline{Hom}}

\DeclareMathOperator{\EXT}{\underline{Ext}}

\newcommand{\Spin}{{\mathop{\text{\textit{Spin}}}\nolimits}}

\newcommand{\Rep}[1]{\ensuremath{\mathbf{\underline{#1}}}}
\newcommand{\barRep}[1]{\ensuremath{\overline{\Rep{#1}}}}
\DeclareMathOperator{\Reg}{Reg}
\DeclareMathOperator{\Ad}{ad}

\newcommand{\textdef}[1]{{\it #1}}

\newcommand{\Xt}{{\ensuremath{\widetilde{X}}}}
\newcommand{\ZZZ}{\ensuremath{{\Z_3\times\Z_3}}}
\newcommand{\Lsheaf}{\ensuremath{\mathscr{L}}}
\newcommand{\Osheaf}{\ensuremath{\mathscr{O}}}
\newcommand{\OsheafXt}{\ensuremath{\mathscr{O}_{\Xt}}}
\newcommand{\OsheafBone}{\ensuremath{\mathscr{O}_{B_1}}}
\newcommand{\OsheafBtwo}{\ensuremath{\mathscr{O}_{B_2}}}
\newcommand{\OsheafP}{\ensuremath{\mathscr{O}_{\CP^1}}}
\newcommand{\Vsheaf}{\ensuremath{\mathscr{V}}}
\newcommand{\Wsheaf}{\ensuremath{\mathscr{W}}}
\newcommand{\Esheaf}{\ensuremath{\mathscr{E}}}
\newcommand{\Fsheaf}{\ensuremath{\mathscr{F}}}
\newcommand{\Hsheaf}{\ensuremath{\mathscr{H}}}
\newcommand{\Hsheafdual}{\ensuremath{\mathscr{H}^\vee}}

\newcommand{\Ssheaf}{\ensuremath{\mathscr{S}}}
\newcommand{\dual}{\ensuremath{\vee}}
\newcommand{\Asheaf}{\ensuremath{\mathscr{A}}}
\newcommand{\Bsheaf}{\ensuremath{\mathscr{B}}}
\newcommand{\Csheaf}{\ensuremath{\mathscr{C}}}
\newcommand{\Qsheaf}{\ensuremath{\mathscr{Q}}}

\newcommand{\LSS}{Leray spectral sequence}

\newenvironment{descriptionlist}{%
\begin{list}%
{}%
{}}%
{\end{list}}

{\everymath{\displaystyle\everymath{}}\array}%
{\endarray}

\begin{document}

\begin{titlepage}
  \begin{flushright}
    hep-th/0505041
    \\
    UPR-1118-T
  \end{flushright}
  \vspace*{\stretch{3}}
  \begin{center}
     \Huge 
     Vector Bundle Extensions, \\
     Sheaf Cohomology,\\
     and the Heterotic Standard Model
  \end{center}
  \vspace*{\stretch{2}}
  \begin{center}
    \begin{minipage}{\textwidth}
      \begin{center}
        \large         
        Volker Braun${}^{1,2}$, 
        Yang-Hui He$^{1}$,
        Burt A. Ovrut${}^{1}$, and
        Tony Pantev${}^{2}$
      \end{center}
    \end{minipage}
  \end{center}
  \vspace*{1mm}
  \begin{center}
    \begin{minipage}{\textwidth}
      \begin{center}
        ${}^{1}$
        Department of Physics,~
        ${}^{2}$
        Department of Mathematics\\        
        David Rittenhouse Laboratory, University of Pennsylvania\\
        209 S. 33rd Street, Philadelphia, PA 19104, USA
      \end{center}
    \end{minipage}
  \end{center}
  \vspace*{\stretch{1}}
  \begin{abstract}
    \normalsize 
    Stable, holomorphic vector bundles are constructed on an torus
    fibered, non-simply connected Calabi-Yau threefold using the
    method of bundle extensions. Since the manifold is multiply
    connected, we work with equivariant bundles on the elliptically
    fibered covering space.  The cohomology groups of the vector
    bundle, which yield the low energy spectrum, are computed using
    the Leray spectral sequence and fit the requirements of particle
    phenomenology. The physical properties of these vacua were
    discussed previously. In this paper, we systematically compute all
    relevant cohomology groups and explicitly prove the existence of
    the necessary vector bundle extensions. All mathematical details
    are explained in a pedagogical way, providing the technical
    framework for constructing heterotic standard model vacua.
  \end{abstract}
  \vspace*{\stretch{5}}
  \begin{minipage}{\textwidth}
    \underline{\hspace{5cm}}
    \centering
    \\
    Email: 
    \texttt{vbraun},
    \texttt{yanghe},
    \texttt{ovrut@physics.upenn.edu};
    \texttt{tpantev@math.upenn.edu}
  \end{minipage}
\end{titlepage}

\newpage

\tableofcontents

\section{Introduction}
\label{sec:intro}

The ultimate goal of string theory is to completely describe the known
forces and particles. While string theory itself is basically unique,
the possible choice of vacua is not. Since low energy physics is
determined by the compactification, the question of whether string
theory has phenomenologically viable vacua is one of the key issues
today. We are not yet able to answer this in full generality.  We are,
however, able to claim an encouraging success. A long-standing problem
in string theory is whether or not one can find compactifications that
produce the correct low energy spectrum, without any exotic matter. By
``exotic'' we mean not only matter fields transforming in
representations that are not in the standard model, but also
additional replicas of quarks and leptons beyond three families. Note
that we are including the right-handed neutrino as a member of each
standard model family, see~\cite{Giedt:2005vx, Langacker:2004xy,
  Fukuda:1998mi}.

To date, most attempts at model building used Type II orientifolds,
see~\cite{Blumenhagen:2000wh, Blumenhagen:2001te, Blumenhagen:2002gw,
  Cvetic:2001tj, Ibanez:2001nd, Cremades:2002qm, Honecker:2004kb,
  Blumenhagen:2005mu, Marchesano:2004yq, Marchesano:2004xz,
  Kokorelis:2004gb, Kokorelis:2002xm, Cvetic:2004ui, Cvetic:2005bn}.
The advantage of this approach is the availability of a conformal
field theory description, in particular Gepner
models~\cite{Blumenhagen:2003su, Blumenhagen:2004cg, Dijkstra:2004ym,
  Dijkstra:2004cc}. This is, however, also the biggest drawback, as
one is always forced to work at special points in the moduli space
with enhanced symmetries and extra massless fields.  Going to more
generic points in this context is very
difficult~\cite{Blumenhagen:2002wn, Blumenhagen:2002vp}. The same
problem is implicit in heterotic orbifold
constructions~\cite{Casas:1988hb, Casas:1989hz, Giedt:2000bi,
  Forste:2004wg, Antoniadis:1987tv, Faraggi:1989ka}.

Instead, our construction employs the $E_8\times E_8$ heterotic
string, both in the strong coupling~\cite{Horava:1995qa,
  Horava:1996ma, Lukas:1997fg, Lukas:1998ew, Lukas:1998hk,
  Lukas:1998tt, Lukas:1998yy} regime of heterotic M-theory and in the
weakly coupled~\cite{Gross:1985fr, Gross:1985rr} regime. Moreover, we
allow for arbitrary vector bundles in the $(0,2)$ model instead of
restricting ourselves to the so-called\footnote{This is really a
  misnomer, there is nothing intrinsically ``standard'' here.}
``standard embedding''~\cite{Greene:1986ar, Greene:1986bm,
  Greene:1986jb, Matsuoka:1986vg, Aspinwall:1987cn}. To find a
realistic, $\Ncal=1$ supersymmetric vacuum of this theory, one needs
to specify a six-dimensional Ricci-flat manifold and an $E_8\times
E_8$ gauge connection satisfying the hermitian Yang-Mills equations.
Fortunately, we do not actually have to solve the equations of motion.
The results of~\cite{MR86h:58038, MR88i:58154} guarantee that any
solution is equivalent to constructing a \CY{} threefold together with
a stable, holomorphic vector bundle. Until now, the standard way to
construct such bundles was to use spectral covers on elliptically
fibered threefolds, see~\cite{Friedman:1997ih, Donagi:1998xe,
  Donagi:1999gc}. However, it turned out to be difficult to construct
realistic matter spectra in this context, see~\cite{Andreas:1999ty,
  DonagiPrincipal, Donagi:1999gc, Curio:2004pf, Andreas:2003zb,
  Diaconescu:1998kg}. Mixing spectral covers with vector bundle
extensions was attempted in~\cite{Donagi:2004ub} for $SU(5)$ bundles,
but failed to yield a phenomenologically viable model.

In this work, we will give a detailed mathematical analysis of the
heterotic standard model that we presented previously
in~\cite{VBphysicsletter, VBmathletter}. For the above reasons, we
will not employ spectral covers to construct vector bundles. Rather,
we use the method of ``bundle extensions'' alone. This method is
discussed in detail, and we give a careful computation of the low
energy spectrum~\cite{Green:1987sp, Green:1987mn, Donagi:2004ia,
  Donagi:2004qk} via the \LSS{}. We have already constructed a
suitable \CY{} threefold in~\cite{dP9Z3Z3} and will take this manifold
as the base space for the necessary vector bundles. This threefold is
torus fibered~\cite{Donagi:2000fw, Donagi:2000zs, Donagi:2000zf,
  Donagi:1999ez, Donagi:2003tb, Ovrut:2003zj, Ovrut:2002jk}, which
gives us good control over the bundles.  By choosing a suitable bundle
and Wilson lines~\cite{Witten:1985xc, Sen:1985eb, Evans:1985vb,
  Breit:1985ud, Breit:1985ns, Ibanez:1986tp}, we are able to find a
compactification which is devoid of any exotic matter fields, except
for an additional Higgs-Higgs conjugate pair. A second Higgs pair is
not ruled out experimentally and may be viewed as a prediction of this
class of models.

\section{Overview}
\label{sec:overview}

The goal of model building is to construct realistic compactifications
of string theory. In this paper, we focus on finding the standard
model with two extra symmetries. Specifically, in addition to the
usual $SU(3)_C\times SU(2)_L\times U(1)_Y$ gauge symmetry, we impose
\begin{itemize}
\item $\Ncal=1$ supersymmetry in $4$ dimensions,
\item an additional $U(1)_{B-L}$ gauge symmetry. This extra symmetry
  naturally suppresses proton decay.
\end{itemize}
We work in the context of the $E_8\times E_8$ heterotic string and
choose the first $E_8$ factor to be in the observable sector. This
factor is then broken down to the desired low energy gauge group. In
principle, there are a number of breaking patterns one could try, but
the minimal pattern in our context is obtained by choosing an
\begin{equation}
  SU(4)\times \ZZZ ~\subset~ E_8
\end{equation}
instanton on the internal \CYm. In other words, we compactify via a
``nonstandard embedding'' rank $4$ gauge bundle together with $\ZZZ$
Wilson lines. In~\cite{dP9Z3Z3}, we constructed a \CY{} threefold $X$
which allows for $\ZZZ$ Wilson lines, and we will review its most
important properties in Section~\ref{sec:CY}.

However, the \CYm{} alone does not determine the heterotic string
compactification. One must, in addition, construct a gauge bundle with
a hermitian Yang-Mills connection. This connection satisfies a
complicated non-linear system of differential equations, but,
fortunately, these can be replaced by an algebraic geometric
criterion, see~\cite{MR86h:58038, MR88i:58154}.  That is, it suffices
to construct a (rank $4$ in our case) stable, holomorphic vector
bundle on $X$. Technically, it is easier to work on the simply
connected covering space $\Xt$, and we will do so throughout this
work. The price one has to pay, however, is that one must construct
equivariant vector bundles on $\Xt$. The precise definition and
relationship of these to vector bundles on $X$ will be developed in
Section~\ref{sec:bundles}, together with some notation.

One way to obtain such bundles is by the so-called ``spectral cover
construction''. While based on a clever trick and exploited thoroughly
in the recent years, this method has always failed to yield vector
bundles for realistic compactifications. Instead, we take the
following starting point. There are two kinds of vector bundles that
one has good control over:
\begin{itemize}
\item Line bundles on the \CY{} threefold $\Xt$.
\item Rank $2$ vector bundles on the base surface of the elliptic
  fibration $\Xt$. 
\end{itemize}
The trivial equation $4=2\cdot 1 + 2\cdot 1$ then suggests that a
certain combination of tensor products would have the desired rank
$4$, and this is precisely the basis for our construction. Of course,
direct sums of vector bundles are never stable. So we have to take
nontrivial extensions.

Once one has constructed a stable, holomorphic bundle one can then
proceed to determine the low energy particle spectrum in the heterotic
compactification. By a standard identification, this is determined by
the sheaf cohomology groups of some associated holomorphic vector
bundles. Computing these requisite cohomology groups is going to be
the main part of this work.

Since there are considerable technical difficulties, we start by
constructing a rank $2$ instanton in the hidden $E_8$ sector. As it
turns out, this is needed for heterotic anomaly cancellation in the
presence of five-branes later on.  It also serves as a simple
introduction to our technology. This will be the subject of
Section~\ref{sec:hiddenE8}. Since the \CYm{} together with the vector
bundle completely determine the compactification, we are then in a
position to read off the low energy spectrum. By a standard
identification between zero modes of the Dirac operator and sheaf
cohomology groups, see~\cite{Green:1987sp, Green:1987mn}, this again
reduces to a question in algebraic geometry. We first apply this to
the hidden gauge bundle in Section~\ref{sec:Hmatter} and conclude
that there are no hidden matter fields.

At this point we are ready to describe the center piece of our work,
the construction of the visible $E_8$ bundle. After a long search, we
found precisely one rank $4$ vector bundle which yields a
phenomenologically viable low energy spectrum. This is described in
Subsection~\ref{sec:SU4bundle}. Our bundle cancels the heterotic
anomaly in a nice way, as we show in
Subsection~\ref{sec:withfivebranes}. In the following two Subsections,
we compute the requisite cohomology groups, check for the existence of
extensions in Subsection~\ref{sec:Vext}, and make sure that the
possible torsion part of the first Chern class vanishes in
Subsection~\ref{sec:c1vanish}. Using these mathematical results, one
can then determine the low energy spectrum and we proceed to do so in
Section~\ref{sec:spectrum}.  One can think of the gauge symmetry
breaking as first taking only the $SU(4)\subset E_8$ instanton into
account, and then subsequently add the effect of the Wilson lines. We
do that in Subsections~\ref{sec:gauge},~\ref{sec:Wilson} and present
the resulting spectrum in~\ref{sec:matter}.

This concludes the main part of our work, but we would still like to
discuss a modification of our vacuum. So far, we have utilized
five-branes in the bulk to cancel the heterotic anomaly. This is only
possible within the context of the strongly coupled heterotic string.
One might ask whether one could perform a small instanton
transition~\cite{Witten:1995gx, Ovrut:2000qi, Donagi:1999jp} and
absorb the five-branes into the hidden $E_8$ bundle and, indeed, this
is possible. Therefore, by a modification of the hidden sector which
we present in Section~\ref{sec:anomaly}, one can work in the weak
coupling regime at the expense of introducing two hidden matter
multiplets. In that case, an $SU(2)\times SU(2)$ bundle is used to
break $E_8$ to $\Spin(12)$, and we find two $\Rep{12}$ matter
multiplets in the hidden $\Spin(12)$.

\section{The Calabi-Yau Manifold}
\label{sec:CY}

\subsection{Fiber Products and Group Actions}
\label{sec:fiberprod}

To compactify the heterotic string so as to preserve $N=1$
supersymmetry, we have to specify two geometric data. First, we must
pick a spacetime background geometry $\R^{3,1}\times X$, where $X$ is
a \CYm. This is what we describe in this Section. Second, we must
construct an $E_8\times E_8$ gauge bundle with a suitable connection.
That will be done in the following Sections.

We take the \CYm{} $X$ to be the space constructed in~\cite{dP9Z3Z3}.
Let us review these results, in as far as we are going to need them
for the remainder of this paper. First of all, $X$ is not simply
connected. Rather,
\begin{equation}
  \pi_1(X)= \ZZZ
  \,.
\end{equation}
This means that there is another \CYm{} $\Xt$ whose $G\eqdef\ZZZ$
quotient is $X$. That is,
\begin{equation}
  X \eqdef 
  \Xt / G = 
  \Xt \Big/ \left( \ZZZ \right)
  \,.
\end{equation}
In~\cite{dP9Z3Z3}, we constructed $\Xt$ as a fiber product of two
$dP_9$ surfaces and then showed that for special values of the moduli
there is a discrete $\ZZZ$ symmetry.

The fiber product $B_1 \times_{\CP^1} B_2$ of two elliptic del Pezzo
surfaces $B_1$ and $B_2$ is defined as follows. We already have
fibrations $\beta_i:B_i\to\CP^1$ such that the fiber over a generic
point $x\in \CP^1$ is a smooth elliptic curve,
\begin{equation}
  \beta_1^{-1}(x) \simeq T^2 \simeq \beta_2^{-1}(x)
  \,.
\end{equation}
The fiber product is the fibration over $\CP^1$ with fiber
$\beta_1^{-1}(x) \times \beta_2^{-1}(x)$, $x\in \CP^1$. Over a generic
point, the fiber is a smooth Abelian surface (a $T^4$). Note that the
base is (complex) one-dimensional and the fiber is two-dimensional, so
we constructed a threefold as desired. If the singular fibers of
$\beta_1$ and $\beta_2$ do not collide then the fiber product $B_1
\times_{\CP^1} B_2$ is again a smooth variety.

To summarize, the \CYm{} $\Xt$ comes with the following chains of
fibrations
\begin{equation}
  \label{eq:projections}
  \vcenter{\xymatrix@!0@=12mm{
      \dim_\C=3: && & \Xt \ar[dr]^{\pi_2} \ar[dl]_{\pi_1} \\
      \dim_\C=2: && B_1 \ar[dr]_{\beta_1} & & 
        B_2 \ar[dl]^{\beta_2} \\
      \dim_\C=1: && & \CP^1 \ar[d]_{\pi} \\
      \dim_\C=0: && & \ptset
      \,.
  }}
\end{equation}
The maps $\pi_1$, $\pi_2$, $\beta_1$, and $\beta_2$ are elliptic
fibrations, and $\pi$ is trivially a $\CP^1$ fibration.

The Hodge diamond of the \CYm{} $X$ is given by
\begin{equation}
  \vcenter{\xymatrix@!0@=7mm@ur{
    1 &  0 &  0 & 1 \\
    0 &  3 &  3 & 0 \\
    0 &  3 &  3 & 0 \\
    1 &  0 &  0 & 1 
  }}
  \,.
\end{equation}

\subsection{Homology Ring}
\label{sec:homology}

Any $dP_9$ surface $B$ has $H_2(B,\Z)=\Z^{10}$. In~\cite{dP9Z3Z3}, we
restricted ourselves to $dP_9$ surfaces with $3I_1$ and $3I_3$
singular fibers. In that case, we defined three special rational
curves $\CP^1\subset B$:
\begin{itemize}
\item The $0$-section $\sigma$ of the elliptic fibration
  $\beta:B\to\CP^1$.
\item The section $\eta$, which generates the torsion part of the
  \MWgrp. 
\item A section $\xi$, which, together with its $\ZZZ$ images,
  generates the remainder of the \MWgrp{}.
\end{itemize}
It turned out that the $\ZZZ$-invariant part of the homology group has
rank two.
\begin{equation}
  H_2(B,\Z)^\ZZZ = \Z f \oplus \Z t
  \,,
\end{equation}
where $f$ is the class of a fiber of the elliptic fibration and $t$ is
the homology sum of three sections\footnote{Here, $\alpha_B$ is a
  $\Z_3$ action related to the overall $\ZZZ$ action. And $\boxplus$
  denotes the addition in the \MWgrp{}, that is, addition of sections
  by point-wise addition in each fiber.}
\begin{equation}
  t = \xi + \alpha_B \xi + (\eta \boxplus \xi)
  \,.
\end{equation}
By definition a section intersects a fiber in a point. Hence, $t$ and
$f$ intersect in $3$ points,
\begin{equation}
  f t = 
  3 \ptset = 
  3 t^2
  \,.
\end{equation}
The $\ZZZ$ invariant part of the homology ring is therefore
\begin{equation}
  \label{eq:Bring}
  H_\ast(B,\Q)^\ZZZ = \Q[f,t] \Big/ \!
  \left< f^2, ft = 3 t^2 \right>
  \,.
\end{equation}
Now, let us return to the \CYm{} $\Xt$. Its $\ZZZ$-invariant divisors
are the pullbacks of the invariant divisors\footnote{We denote the
  invariant divisors on $B_1$ and $B_2$ by $f$ and $t$ and elect to
  not index them separately. It will always be clear from the context
  which surface we are referring to.} on $B_1$ and $B_2$, which we
label as
\begin{equation}
\begin{split}
  \tau_1 ~\eqdef&~ \pi_1^{-1}(t) \\
  \tau_2 ~\eqdef&~ \pi_2^{-1}(t) \\
  \phi   ~\eqdef&~ \pi_1^{-1}(f) = \pi_2^{-1}(f)
  \,.
\end{split}
\end{equation}
The intersection numbers on $\Xt$ then follow from
eq.~\eqref{eq:Bring}. We find that the invariant part of the homology
groups in even degrees is
\begin{equation}
  \label{eq:intersectionRing}
  H_\even\left(\Xt,\Q\right)^\ZZZ=
  \Q[\phi,\tau_1,\tau_2] \Big/\!
  \left< 
    \phi^2,\, 
    \phi\tau_1 = 3 \tau_1^2 ,\,
    \phi\tau_2 = 3 \tau_2^2
  \right>
  \,.
\end{equation}
For practical purposes it is useful to switch to a Gr\"obner basis,
for example with the lexicographic term ordering $\phi \succ \tau_1
\succ \tau_2$,
\begin{equation}
  \left< 
    \phi^2,\, 
    \phi\tau_1 = 3 \tau_1^2 ,\,
    \phi\tau_2 = 3 \tau_2^2
  \right>
  =
  \left< 
    \phi^2 ,\,
    \phi \tau_1 = 3 \tau_1^2 ,\,
    \phi \tau_2 = 3 \tau_2^2 ,\,
    \tau_1^3 ,\,
    \tau_1^2 \tau_2 = \tau_1 \tau_2^2 ,\,
    \tau_2^3
  \right>
  \,.
\end{equation}
Then one can easily bring any polynomial in $\phi$, $\tau_1$, $\tau_2$
into the standard form
\begin{equation}
  \begin{aligned}
    H_0\left(\Xt,\Q\right)^\ZZZ 
    \simeq
    H^6\left(\Xt,\Q\right)^\ZZZ 
    &=~
    \Q \tau_1 \tau_2^2
    \,,
    \\
    H_2\left(\Xt,\Q\right)^\ZZZ 
    \simeq
    H^4\left(\Xt,\Q\right)^\ZZZ 
    &=~
    \Q \tau_1^2        \oplus  
    \Q \tau_1 \tau_2   \oplus  
    \Q \tau_2^2
    \,,
    \\
    H_4\left(\Xt,\Q\right)^\ZZZ 
    \simeq
    H^2\left(\Xt,\Q\right)^\ZZZ 
    &=~
    \Q \phi           \oplus 
    \Q \tau_1          \oplus  
    \Q \tau_2 
    \,,
    \\
    H_6\left(\Xt,\Q\right)^\ZZZ 
    \simeq
    H^0\left(\Xt,\Q\right)^\ZZZ 
    &=~
    \Q    
    \,.
  \end{aligned}
\end{equation}
Note that the natural generator of $H^6\big(\Xt,\Q\big)^\ZZZ$ is not
primitive, that is, it is a multiple of the generator in integral
cohomology. In fact, it is three times the generator of
$H^6\big(\Xt,\Z\big)^\ZZZ$,
\begin{equation}
  \tau_1^2 \tau_2 = \tau_1 \tau_2^2 = 3 \ptset
  \,.
\end{equation}
Getting the normalization correct is, of course, important for index
computations in the following.

\section{Notation and Conventions for Bundles}
\label{sec:bundles}

\subsection{Line Bundles, Ideal Sheaves, and Extensions}
\label{sec:bundleintro}

Having studied the \CY{} threefold $\Xt$, we now want to construct
vector bundles. We start with some basics, which will help to set the
notation in the remainder of this work.

All vector bundles that we consider are holomorphic, that is, the
defining transition functions are holomorphic. The simplest vector
bundle is just the trivial line bundle $\Xt\times\C\rightarrow \Xt$.
The sections of the trivial line bundle are simply holomorphic
functions $f:\Xt\to\C$, and any such function must be constant since
$\Xt$ is compact. But on each coordinate chart $U\subset \Xt$, there
are many holomorphic functions. So while considering holomorphic
functions that are defined everywhere on $\Xt$ is unenlightening, the
holomorphic functions on open subsets can be interesting. This is why
one works with the sheaf of holomorphic functions, $\OsheafXt$, which
assigns to each open set $U\subset \Xt$ the holomorphic functions on
$U$.

Now, technically, $\Xt\times \C$ is the trivial line bundle and
$\OsheafXt$ is the sheaf of local sections of the trivial line bundle.
We will not make that distinction in the following and use either to
denote the line bundle.

Just like the sheaf of local holomorphic functions, one can define the
sheaf of local holomorphic functions which vanish at some points. This
is called the ideal sheaf of the set of points. In particular, we will
use an ideal sheaf on the surface $B_2$ in the following. We write
$I_k$ for the functions on $B_2$ vanishing at a giving set of $k$
points. Note that an ideal sheaf on a surface is not quite a vector
bundle, but, rather, it contains ``point defects''.

Coming back to bundles, there is a simple way to describe all line
bundles on any variety $Y$ using the correspondence
\begin{equation*}
  \Big\{ 
  \text{Divisors } D \Big\} \Big/ \!
  \sim
  \quad
  \stackrel{1:1}{\longleftrightarrow}
  \quad  
  \Big\{ 
  \text{Line bundles } \Osheaf_Y(D) 
  \Big\}
\end{equation*}
between divisors and line bundles. On $\Xt$, $B_1$, $B_2$, $\CP^1$ the
``linear equivalence'' relation $\sim$ just amounts to taking the
homology class of the divisor. Hence, every line bundle is of the form
\begin{itemize}
\item $\OsheafXt(x_1 \tau_1 + x_2 \tau_2 + x_3 \phi)$, 
  $x_1,x_2,x_3 \in \Z$.
\item $\Osheaf_{B_i}(y_1 t + y_2 f)$, $y_1,y_2 \in \Z$.
\item $\OsheafP(n)\eqdef\OsheafP\big(n \ptset\big)$, $n \in \Z$.
\end{itemize}

\subsection{Equivariant Structures}
\label{sec:bundleequivariant}

Our ultimate goal is, of course, to construct holomorphic vector
bundles on the quotient \CY{} threefold $X=\Xt/\big(\ZZZ\big)$. But,
unfortunately, vector bundles on $X$ and vector bundles on $\Xt$ are
only distantly related. Really we want to exploit the one-to-one
correspondence
\begin{equation}
  \left\{
    \begin{array}{c}
      \text{Vector bundles} \\
      \text{on}~X=\Xt\Big/\big(\ZZZ\big)
    \end{array}
  \right\}
  \quad
  \stackrel{1:1}{\longleftrightarrow}
  \quad
  \left\{
    \begin{array}{c}
      \ZZZ~\text{Equivariant vector} \\
      \text{bundles on}~\Xt
    \end{array}
  \right\}
\end{equation}
Let us pause to define equivariant vector bundles. First of all, just
as vector bundles are defined over a fixed base space, equivariant
vector bundles are defined over a fixed $G$-space, that is, a
topological space with an action of the group $G$. An equivariant
vector bundle is then a pair $(\Esheaf, \phi)$ consisting of an
ordinary vector bundle $\Esheaf$ together with an action of the group
$\phi_g:\Esheaf\to \Esheaf$, $g\in G$. It is crucial that the $|G|$
maps $\phi_g$ represent the group, that is $\phi_g \phi_{g'} = \phi_{g
  g'}$.  Finally, the group action on the vector bundle must cover the
action on the base space, that is, map the fiber $\Esheaf_p$ over the
point $p$ to the fiber over the image point $g(p)$, in other words
$\phi_g(\Esheaf_p) = \Esheaf_{g(p)}$.

An important point is that this group action is not unique. Consider
(the sheaf of sections of) the trivial line bundle $\OsheafXt$ on the
\CY{} threefold $\Xt$. It is clearly invariant under the $\ZZZ$
action, but there are different choices for how the group acts on
$\OsheafXt$. To illustrate this, let us look at a single $\Z_3$ action
$g:\Xt\to\Xt$. Now a $\Z_3$ equivariant structure on $\OsheafXt$ is a
map $\gamma:\OsheafXt\to\OsheafXt$ covering $g$. That is, $\gamma$
maps elements in the vector space over a point to the vector space
over the $g$-image of that point. In other words, the diagram
\begin{equation}
  \label{eq:equivariant}
  \vcenter{\xymatrix{
    \OsheafXt \ar[d] \ar[r]_{\gamma} & \OsheafXt \ar[d] \\
    \Xt \ar[r]_{g} & \Xt 
  }}
\end{equation}
commutes. There is an obvious such map $\gamma$: identify
$\OsheafXt\simeq \Xt\times \C$ and let $\gamma$ not act on the vector
space at all. That is,
\begin{equation}
  \gamma : \Xt\times \C \to \Xt\times \C,~
  (p,v) \mapsto \big( g(p), v \big)
  \,.
\end{equation}
But this is not the only choice, and we could combine it with any
third root of unity multiplying the vector component. In other words,
for any character\footnote{A character of a group $G$ is a
  homomorphism $\chi:G\to \C^\times$. Since the group is finite in our
  case, it is actually a map $G\to U(1)$. Note that this is not quite
  the same as the character of a representation, the latter being the
  traces of representation matrices.} $\chi$ of $\Z_3$, there is
another equivariant structure
\begin{equation}
  \chi \gamma: \Xt\times \C \to \Xt\times \C,~
  (p,v) \mapsto \Big( g(p), \chi(g) v \Big)
  \,.
\end{equation}

This is why, in the following, we need a notation to express the
equivariant structure on a line bundle. We fix generators $g_1$ and
$g_2$ of the $\ZZZ$ group,
\begin{equation}
  G = 
  G_1 \times G_2 =
  \{e,g_1,g_1^2\} \times \{e,g_2,g_2^2\}
  \simeq \ZZZ
  \,,
\end{equation}
and choose the following generators of the character ring, 
\begin{equation}
\begin{aligned}
  \chi_1(g_1) &= \omega 
  & \qquad
  \chi_1(g_2) &= 1
  \\
  \chi_2(g_1) &= 1
  &
  \chi_2(g_2) &= \omega 
  \,,
\end{aligned}
\end{equation}
where $\omega = e^{\frac{2\pi i}{3}}$.  We then always take the $G$
action on the trivial line bundle $\Osheaf_Y$ to be the pure
translation, and write $\chi\Osheaf_Y$ for the translation composed
with multiplication by a character,
\begin{equation}
  g:  
  \chi \Osheaf_Y \to 
  \chi \Osheaf_Y
  \,,~
  \big( p,v \big) \mapsto 
  \Big( g_i(p), \chi(g) v\Big)
  \,.
\end{equation}
This uniquely determines the action on trivial bundles. In addition,
we need to consider vertical bundles, that is, line bundles whose
associated divisor is a multiple of the elliptic fiber. For that, we
restrict to one of the two $G$ fixed points on the base $\CP^1$. So
pick once and for all
\begin{equation}
  0 \in \CP^1,\quad G\cdot 0=0
  \,.
\end{equation}
Then any vertical bundle $\Osheaf_{B_i}(nf)$ restricts to a trivial
bundle on $f=\beta_i^{-1}(0)$ and, hence, is of the form
\begin{equation}
  \Osheaf_{B_i}(nf) \big|_{f=\beta_i^{-1}(0)} = 
  \chi \Osheaf_f
  \,.
\end{equation}
We then label the equivariant structure of $\Osheaf_{B_i}(nf)$ by the
character $\chi$, that is,
\begin{equation}
  \Lsheaf = \chi \Osheaf_{B_i}(nf)
  \quad \stackrel{\mathrm{def}}{\Leftrightarrow} \quad
  \Lsheaf \simeq \Osheaf_{B_i}(nf)
  \quad\text{and}\quad
  \Lsheaf \big|_{f=\beta_i^{-1}(0)} = 
  \chi \Osheaf_f  
  \,.
\end{equation}
In the same way, we label the equivariant structure on line bundles on
$\CP^1$ and of vertical line bundles on $\Xt$ as
\begin{align}
  \Lsheaf = \chi \OsheafP(n)
  \quad &\stackrel{\mathrm{def}}{\Leftrightarrow} \quad
  \Lsheaf \simeq \OsheafP(n)
  \quad\text{and}\quad
  \Lsheaf \big|_0 = 
  \chi 
  \,,
  \\
  \Lsheaf = \chi \OsheafXt(n\phi)
  \quad &\stackrel{\mathrm{def}}{\Leftrightarrow} \quad
  \Lsheaf \simeq \OsheafXt(n\phi)
  \quad\text{and}\quad
  \Lsheaf \big|_{\phi=(\beta_i\circ\pi_i)^{-1}(0)} = 
  \chi \Osheaf_\phi
  \,.
\end{align}
Here we also identified the one-dimensional
representation of $G$ determined by $\chi$ with the character
$\chi:G\to U(1)$. This abuse of notation will be continued throughout
this paper.

\subsection{Equivariant vs. Invariant Bundles}
\label{sec:invequiv}

Our insistence on explicitly denoting the group action on every bundle
is important. Most vector bundles on $X$ do not admit a $\ZZZ$ action,
even if their Chern classes are invariant, and do not correspond to
vector bundles on the quotient $X/G$. This is why we must be very
careful to construct equivariant vector bundles.

The underlying problem can be made very explicit. Consider the line
bundle $\OsheafXt(\tau_1)$, one of the simplest invariant line bundles
one can possibly write down. Yet $\OsheafXt(\tau_1)$ does not have any
$\ZZZ$ action, and cannot be made into an equivariant line bundle.
This can be seen as follows. The elliptic fibration $\pi_2:\Xt\to B_1$
has one elliptic fiber which is mapped to itself under the $\ZZZ$
action and, moreover, on which both generators act as translation by
an order $3$ point. Call this elliptic fiber $E\simeq \C/\Lambda$. The
divisor $\tau_1$ intersects this elliptic curve in three points
\begin{equation}
  \tau_1 \cap E = \{P_1, P_2, P_3\}
  \,,
\end{equation}
and, hence, the restriction of the line bundle $\OsheafXt(\tau_1)$ to
$E$ is
\begin{equation}
  \OsheafXt(\tau_1)\big|_E = \Osheaf_E(P_1+P_2+P_3)
  \,.
\end{equation}
Now the moduli space of line bundles on an elliptic curve (the Picard
variety) looks like 
\begin{equation}
  \Pic(E) \simeq \Z \times T^2
  \,.
\end{equation}
In other words, the line bundles are determined by one integer (the
first Chern class) and one complex number. The latter is a continuous
modulus, which is just the sum of the coordinates (modulo $\Lambda$)
of the points of the corresponding divisor. That is, in our case
\begin{equation}
  P_1 \boxplus P_2 \boxplus P_3 \in E
  \,.
\end{equation}
Obviously, if we pick an order $3$ point and translate $P_1$, $P_2$,
and $P_3$ by it then the sum does not change (again, modulo
$\Lambda$). Hence $g^\ast \Osheaf_E(P_1+P_2+P_3) \simeq
\Osheaf_E(P_1+P_2+P_3)$ for each $g\in \ZZZ$, and we can pick one such
map
\begin{equation}
  \phi_{g_i}: ~
  \Osheaf_E(P_1+P_2+P_3)
  ~\stackrel{\sim}{\longrightarrow}~
  \Osheaf_E(P_1+P_2+P_3)
\end{equation}
for each of the two generators $g_1, g_2 \in \ZZZ$.  Now one might
think that these maps would turn $\Osheaf_E(P_1+P_2+P_3)$ into an
equivariant line bundle, and it is indeed $\Z_3$ equivariant for the
$\Z_3$ subgroups generated by $g_1$ or $g_2$. But one cannot turn it
into a $\ZZZ$ equivariant line bundle, simply because its first Chern
class
\begin{equation}
  c_1\big(\Osheaf_E(P_1+P_2+P_3)\big)=3  
\end{equation}
is not divisible by $|\ZZZ|=9$. The maps $\phi_{g_1}$, $\phi_{g_2}$
fail to define an equivariant line bundle because they do not commute,
whereas $g_1$, $g_2$ of course do commute. At most, one can choose 
them to commute up to a third root of unity $\omega=e^\frac{2\pi
  i}{3}$, 
\begin{equation}
  \phi_{g_1} \circ \phi_{g_2} = \omega\, \phi_{g_2} \circ \phi_{g_1}
  \,.
\end{equation}
Put differently, the line bundle $\Osheaf_E(P_1+P_2+P_3)$ can only be
equivariant under the Heisenberg group $G_H$, that is, the central
extension
\begin{equation}
  0
  \longrightarrow
  \Z_3
  \longrightarrow
  G_H
  \longrightarrow
  \ZZZ
  \longrightarrow
  0
  \,.
\end{equation}
Since we will make use of it in the following, let us mention an
elementary fact from the representation theory of the Heisenberg
group. There is only one irreducible representation such that the
central $\Z_3$ acts by multiplication with $\omega$. This
representation is three dimensional. In terms of matrices it is
generated by
\begin{equation}
  \rho(g_1) = 
  \begin{pmatrix}
    0 & 1 & 0 \\
    0 & 0 & 1 \\
    1 & 0 & 0 
  \end{pmatrix}
  \,,\qquad
  \rho(g_2) = 
  \begin{pmatrix}
    1 & 0 & 0 \\
    0 & \omega & 0 \\
    0 & 0 & \omega^2 
  \end{pmatrix}
  \,.
\end{equation}

\subsection{Canonical Bundles}
\label{sec:K}

One basic and technique for computing cohomology groups is to use
Serre duality. For any variety $Y$, it relates the cohomology groups
\begin{equation}
  \label{eq:serre}
  H^{\dim_\C(Y)-i}\Big( Y,\, \Fsheaf\Big)^\dual =
  H^{i}\Big( Y,\, \Fsheaf^\dual \otimes K_Y \Big)
\end{equation}
if $\Fsheaf$ is vector bundle and
\begin{equation}
  \label{eq:serreExt}
  H^{\dim_\C(Y)-i}\Big( Y,\, \Fsheaf\Big)^\dual =
  \Ext^{i}\Big( \Fsheaf, K_Y \Big)
\end{equation}
for arbitrary sheaves $\Fsheaf$. Here, $K_Y$ is the canonical bundle
of $Y$, that is, the sheaf of $\dim_\C(Y)$-forms. The duality follows
from a perfect pairing defined by integrating over the manifold $Y$.
Hence the appearance of the canonical bundle, that is, the line bundle
of top dimensional holomorphic differentials.

More important for our purposes, there is a relative (fiberwise)
version which, for any elliptic fibration $\pi:Y\to Z$, yields
\begin{equation}
  \label{eq:relativeduality}
  \pi_\ast \big(\Fsheaf^\vee\big) =
  \Big( R^1\pi_\ast \big( \Fsheaf \otimes K_{Y|Z}\big)  \Big)^\vee  
  \,,
\end{equation}
where $K_{Y|Z}$ denotes the relative canonical bundle
\begin{equation}
  K_{Y|Z} \eqdef
  K_Y \otimes \pi^\ast K_Z^\vee
  \,.
\end{equation}
To make use of these dualities, we have to know how the canonical
bundles transform under the $G=G_1\times G_2\simeq \ZZZ$ action. It is
well known that, up to the character coming from the group action,
\begin{equation}
  \label{eq:canonicalprelim}
  K_\Xt \simeq \OsheafXt
  \,,\quad
  K_{B_i} \simeq \Osheaf_{B_i}(-f)
  \,,\quad
  K_{\CP^1} \simeq \OsheafP(-2)
  \,.
\end{equation}
To determine the extra phase, all one has to do is to look at the
behavior in one of the two $G$-stable fibers. Without loss of
generality, we restrict our attention to $E_i=\beta_i^{-1}(0)$. From
the analysis of the $G$ action in~\cite{dP9Z3Z3}, we know that $G_1$
acts as an order $3$ translation on one of the fibers and complex
multiplication on the other fiber. Again without loss of generality,
we assume that $G_1$ acts via complex multiplication on $E_1$, thereby
fixing the complex structure modulus of that elliptic curve. The
complex structure on $E_2$, on the other hand, remains unconstrained
and is one of the moduli of the \CY{} threefold. To summarize, the
fibers over $0\in\CP^1$ are as depicted in Figure~\ref{fig:fibers}.
\begin{figure}[htbp]
  \centering
  \input{torusfibers.pstex_t}
  \caption{Fibers over $0\in \CP^1$.}
  \label{fig:fibers}
\end{figure}

The $G$ action can easily be written down in terms of local
coordinates $(z,u,v)$. First, $G_1$ acts as rotation on the base
$\CP^1$. Therefore, in terms of the local coordinate around the fixed
point it acts by
\begin{equation}
  \label{eq:zaction}
  g_1: z\mapsto \omega z 
  \,, \qquad
  g_2: z\mapsto z
  \,,
\end{equation}
where $\omega$ is a third root of unity. Which, without loss of
generality, can be taken to be $\omega \eqdef e^{\frac{2\pi i}{3}}$.
Second, consider the $G$ action on the elliptic curve $E_2$. Since its
complex structure $\lambda_2$ 
is arbitrary, each generator of $G$ must act by translation. Hence,
the action must be
\begin{equation}
  \label{eq:vaction}
  g_1: v\mapsto v + \frac{1}{3} 
  \,,\qquad
  g_2: v\mapsto v + \frac{\lambda_2}{3}
  \,.
\end{equation}
Finally, let us examine the $G$ action on $E_1$. By definition, $g_2$
acts on every fiber as a translation. In contrast, $g_1$ acts by
complex multiplication on the elliptic curve. Moreover, this phase is
coupled to the phase in the transformation of the coordinate $z$ on
the base $\CP^1$, eq.~\eqref{eq:zaction}, since the holomorphic
volume form $\Omega \sim \diff u\, \diff v \, \diff z$ must be
invariant under the $G$ action. Therefore, the $G$ action on the $u$
coordinate is as follows,
\begin{equation}
  \label{eq:uaction}
  g_1: u\mapsto \omega^2 u 
  \,, \qquad
  g_2: u\mapsto u + \frac{1}{3}
  \,.
\end{equation}
Now we know the action on the local coordinates $u$, $v$, and $z$.
Hence, we also know the $G$ action on the top dimensional holomorphic
differentials, that is, the local sections of the canonical bundle.
This fixes the missing characters in eq.~\eqref{eq:canonicalprelim} to
be
\begin{equation}
  \label{eq:CanonicalBundles}
  \vcenter{\xymatrix@!0@C=24mm@R=32mm{
      \dim_\C=3: && & 
      \Big(\Xt,\, K_{\Xt}= \OsheafXt\Big)
      \ar[dr]_{\pi_2}^
      {\displaystyle K_{\Xt|B_2}=\chi_1^2\OsheafXt(\phi)}
      \ar[dl]^{\pi_1}_
      {\displaystyle K_{\Xt|B_1}=\OsheafXt(\phi)}
      \\
      \dim_\C=2: && 
      \Big(B_1,\, K_{B_1}= \OsheafBone(-f) \Big)
      \ar[dr]^{\beta_1}_
      {\displaystyle K_{B_1|\CP^1}=\chi_1^2\OsheafBone(f)} & & 
      \Big(B_2,\, K_{B_2}=\chi_1 \OsheafBtwo(-f) \Big)
      \ar[dl]_{\beta_2}^
      {\displaystyle K_{B_2|\CP^1}=\OsheafBtwo(f)} \\
      \dim_\C=1: && & 
      \Big(\CP^1,\, K_{\CP^1} = \chi_1 \OsheafP(-2) \Big)
      \,.
  }}
\end{equation}

\section{\texorpdfstring%
{The Hidden $\mathbf{E_8}$ Bundle}%
{The Hidden E8 Bundle}}
\label{sec:hiddenE8}

\subsection{Constructing Vector Bundles by Extension}
\label{sec:extensionH}

For simplicity, we start by constructing an $SU(2)$ instanton on the
hidden brane. This allows us to introduce the techniques we are going
to use later on in a simpler setting. 

We define $\Hsheaf$ to be an extension of the line bundle
$\OsheafXt(-2\tau_1 - \tau_2 + \phi)$ by $\OsheafXt( 2 \tau_1 + \tau_2
- \phi)$. That is, by definition, $\Hsheaf$ is the middle term in a
short exact sequence
\begin{equation}
  \label{eq:Hdef}
  0 
  \longrightarrow
  \OsheafXt( 2 \tau_1 + \tau_2 - \phi)
  \longrightarrow
  \Hsheaf
  \longrightarrow
  \OsheafXt(-2\tau_1 - \tau_2 + \phi)  
  \longrightarrow
  0
  \,.
\end{equation}
The first Chern classes of the line bundles obviously add up to zero.
Hence, we really obtain an $SU(2)$ rather than just a $U(2)$ bundle.

We furthermore demand that the extension be generic, ruling out the
(slope-unstable) direct sum $\OsheafXt( 2 \tau_1 + \tau_2 -
\phi)\oplus \OsheafXt(-2\tau_1 - \tau_2 + \phi)$. Apart from
disallowing special cases, we are not going to impose any further
restrictions on the extensions. That is, we are not constraining the
vector bundle moduli to specific values.

Of course, this is only possible if non-trivial extensions exist. So
we must compute the possible extensions, and show that there exist
more than the trivial extension in order to justify our assumptions.
The space of extensions is
\begin{equation}
  \Ext^1\Big( 
  \OsheafXt(-2\tau_1 - \tau_2 + \phi)  
  ,~
  \OsheafXt( 2 \tau_1 + \tau_2 - \phi)
  \Big)
  \,.
\end{equation}
However, not every such extension gives rise to an equivariant vector
bundle. The line bundles are equivariant, but if the class of the
extension changes under the group action then we do not obtain a
group action on the extension. Only the $G$-invariant part of the
$\Ext^1$ yields an equivariant vector bundle.

We are going to compute the space of extensions in
Subsection~\ref{sec:Hext}, and find for its $G$-invariant part that
\begin{equation}
  \Ext^1\Big( 
  \OsheafXt(-2\tau_1 - \tau_2 + \phi)  
  ,~
  \OsheafXt( 2 \tau_1 + \tau_2 - \phi)
  \Big)^G = 
  6
  \,.
\end{equation}
Hence, the assumption that a generic (nontrivial) extension of the
form eq.~\eqref{eq:Hdef} exists is justified.

\subsection{Sheaf Cohomology on Elliptic Fibrations}
\label{sec:cohomologyH}

\subsubsection{Cohomology of Line Bundles}

In order to determine the low energy spectrum, we must compute the
cohomology groups of the vector bundle $\Hsheaf$. The general way to
do this is as follows. First, we can exploit the fibration structure,
eq.~\eqref{eq:projections}, and compute the cohomology of line bundles
by successively pushing down all the way to a point. In the second
step, we then determine the cohomology of an extension of line bundles
by the associated long exact sequence in cohomology.

For example, let us start with the line bundle $\OsheafXt( 2 \tau_1 +
\tau_2 - \phi)$. First, we use the elliptic fibration $\pi_1:\Xt\to
B_1$ to relate the cohomology groups $H^i\big(\Xt,\OsheafXt( 2 \tau_1
+ \tau_2 - \phi)\big)$ on the threefold $\Xt$ to cohomology groups on
the complex surface $B_1$. In general, for any fibration, the
cohomology groups on the whole space are determined by a combination
of the cohomology of the fibers and the cohomology of the base. This
is made precise by the \LSS:
\begin{equation}
\label{eq:LerayXt}
\begin{split}
  E_2^{p,q}(\Xt|B_1) \,&=
  H^p\Big( B_1,\,
    R^q\pi_{1\ast} 
    \OsheafXt( 2 \tau_1 + \tau_2 - \phi) 
  \Big)
  \\  &\Rightarrow \quad
  H^{p+q}\Big( \Xt,\, 
    \OsheafXt( 2 \tau_1 + \tau_2 - \phi) 
  \Big)
  \,.
\end{split}
\end{equation}
The derived pushdown $R^q\pi_{1\ast}$ is nothing but the cohomology
along the fiber, that is, some sheaves on $B_1$ that have to be
computed first. Then we must determine their cohomology groups (on the
surface $B_1$). Starting with cohomology on $B_1$, the \LSS{} then
converges (denoted by ``$\Rightarrow$'') to the cohomology on the
threefold $\Xt$. By dimension, the terms of the first quadrant
spectral sequence $E_2^{p,q}(\Xt|B_1)$ vanish for $p>\dim_\C(B_1)=2$
and for $q>\dim_\C(\Xt)-\dim_\C(B_1)=1$.

To completely determine the ingredients in the spectral sequence, we
will use the following two facts.
\begin{itemize}
\item The projection formula, which in general reads
  \begin{equation}
    \label{eq:projectionformula}
    R^q\pi_\ast \Big( \Esheaf \otimes \pi^\ast \Fsheaf \Big)
    = 
    R^q\pi_\ast \big( \Esheaf \big) \otimes \Fsheaf
  \end{equation}
  for any fibration $\pi$, arbitrary sheaf $\Esheaf$, and vector
  bundle $\Fsheaf$. In the case at hand, we conclude that
  \begin{equation}
    \begin{split}
      R^q\pi_{1\ast} 
      \OsheafXt( 2 \tau_1 + \tau_2 - \phi) &=
      R^q\pi_{1\ast} \Big(
      \pi_1^{\ast}\big(\OsheafBone(2t-f)\big) + 
      \pi_2^{\ast}\big(\OsheafBtwo(t)\big) 
      \Big) = 
      \\ &=
      \OsheafBone(2t-f) \otimes
      R^q\pi_{1\ast} \pi_2^{\ast}\big(\OsheafBtwo(t)\big)             
      \,.
    \end{split}
  \end{equation}
\item The commutativity of the projections in
  eq.~\eqref{eq:projections}, which implies
  \begin{equation}
    \begin{split}
      \big(R^q\pi_{1\ast}\big) \circ \big(\pi_2^\ast\big)
      &= \big(\beta_1^\ast\big) \circ \big(R^q \beta_{2\ast}\big)
      \,,
      \\
      \big(R^q\pi_{2\ast}\big) \circ \big(\pi_1^\ast\big)
      &= \big(\beta_2^\ast\big) \circ \big(R^q \beta_{1\ast}\big)
      \,.      
    \end{split}    
  \end{equation}
\end{itemize}
Using these, the starting point of the spectral sequence
eq.~\eqref{eq:LerayXt}, is completely determined in terms of line
bundles on $B_1$ alone,
\begin{equation}
  \begin{split}
    &
    H^p\Big( B_1,\,
    R^q\pi_{1\ast} 
    \OsheafXt( 2 \tau_1 + \tau_2 - \phi)   
    \Big)
    = \\ &\qquad=
    H^p\Big( B_1,\,
    \OsheafBone(2t-f) \otimes
    R^q\pi_{1\ast} \pi_2^{\ast}\big(\OsheafBtwo(t)\big)             
    \Big)
    = \\ &\qquad=
    H^p\Big( B_1,\,
    \OsheafBone(2t-f) \otimes
    \beta_1^{\ast} R^q\beta_{2\ast}\big(\OsheafBtwo(t)\big)             
    \Big)        
    \,.
  \end{split}  
\end{equation}
It remains to compute the cohomology groups on $B_1$. We again proceed
by pushing down one step to the base $\CP^1$. But now we end up with
two spectral sequences, corresponding to the case $q=0$ and $q=1$ in
the previous equation. In the following, we will describe the $q=0$
case, the computation for $q=1$ being completely analogous.

The \LSS{} now reads
\begin{equation}
\label{eq:Leray2}
\begin{split}
  E_2^{p,q}(B_1|\CP^1) ~=&~ 
  H^p\bigg( \CP^1,\,
    R^q\beta_{1\ast} 
    \Big[
      \OsheafBone(2t-f) \otimes
      \beta_1^{\ast} \beta_{2\ast}\OsheafBtwo(t)
    \Big]
  \bigg)
  = \\ ~=&~
  H^p\Big( \CP^1,\,
    R^q\beta_{1\ast} \OsheafBone(2t-f) 
    \otimes
    \beta_{2\ast}\OsheafBtwo(t)
  \Big)
  = \\ ~=&~
  H^p\Big( \CP^1,\,
    R^q\beta_{1\ast} \OsheafBone(2t) 
    \otimes
    \beta_{2\ast}\OsheafBtwo(t)
    \otimes 
    \OsheafP(-1)
  \Big)
  \\
  \Rightarrow& \quad
  H^{p+q}\Big( B_1,\, 
    \pi_{1\ast} \OsheafXt( 2 \tau_1 + \tau_2 - \phi) 
  \Big)    
  \,.
\end{split}
\end{equation}
The pushdown of the line bundles on $B_1$ can be found by a
straightforward application of the long exact sequence for pushdowns.
We defer the details to Appendix~\ref{sec:pushdownformula}, here only
listing the result for convenience.
\begin{subequations}
\begin{gather}
  \label{eq:Bpushdownformulasummary}
  \Osheaf_{B_i}(nf) = \beta_i^\ast \OsheafP(n)
  \,, \quad n\in \Z
  \\
  \begin{align}
    \beta_{i\ast} \Osheaf_{B_i}(2t)  &= 6 \OsheafP
    &
    \beta_{i\ast} \Osheaf_{B_i}(-2t) &= 0
    &
    R^1\beta_{i\ast} \Osheaf_{B_i}(2t)  &= 0  
    \\
    \beta_{i\ast} \Osheaf_{B_i}(t)   &= 3 \OsheafP
    &
    \beta_{i\ast} \Osheaf_{B_i}(-t)  &= 0
    &
    R^1\beta_{i\ast} \Osheaf_{B_i}(t)   &= 0
  \end{align}
  \\
  \begin{align}
    R^1\beta_{1\ast} \OsheafBone(-t)  &= 3 \chi_1 \OsheafP(-1)
    &
    R^1\beta_{1\ast} \OsheafBone(-2t) &= 6 \chi_1 \OsheafP(-1)
    \\
    R^1\beta_{2\ast} \OsheafBtwo(-t)  &= 3 \OsheafP(-1)
    &
    R^1\beta_{2\ast} \OsheafBtwo(-2t) &= 6 \OsheafP(-1)
  \end{align}
\end{gather}
\end{subequations}
Hence, the $E_2$ term of the spectral sequence is 
\begin{equation}
  E_2^{p,q}(B_1|\CP^1) =
  \begin{cases}
    0 & q=1 \\
    H^p\big(\CP^1,\, 18 \OsheafP(-1) \big)& q=0 
    \,.
  \end{cases}
\end{equation}
Sheaves on $\CP^1$ are particularly simple. All vector bundles split
into the sum of line bundles. Furthermore, the global sections of a
line bundle are polynomials of the given degree. The number of
monomials and their transformation under the $\ZZZ$ action can be
found easily, determining the zeroth cohomology group. The first
cohomology group is then Serre dual to the zeroth cohomology group. It
follows that
\begin{equation}
\label{eq:P1cohomology}
\begin{split}
  H^0\big(\CP^1,\, \chi \OsheafP(n) \big) =&\,
  \begin{cases}
    0\,,   & n<0 \\
    \chi \sum_{i=0}^n \chi_1^i\,, & n\geq 0
  \end{cases}
  \\
  H^1\big(\CP^1,\, \chi \OsheafP(n) \big) =&\,
  H^0\Big(\CP^1,\, 
  \big( \chi \OsheafP(n) \big)^\vee \otimes K_{\CP^1} \Big)^\vee
  \,.
\end{split}
\end{equation}
Putting everything together, we see that all cohomology groups vanish
and, hence, $E_2^{p,q}\big(\Xt|B_1\big)=0$. There is no possibility for
nonvanishing differentials and, therefore, all cohomology groups of the
line bundle $\OsheafXt( 2 \tau_1 + \tau_2 - \phi)$ actually vanish,
\begin{equation}
  H^\ast\Big( \Xt, \,
    \OsheafXt( 2 \tau_1 + \tau_2 - \phi) 
  \Big) = 0
  \,.
\end{equation}
Similarly, one can compute the cohomology of the dual line bundle or
simply invoke Serre duality. In either case, one finds
\begin{equation}
  H^\ast\Big( \Xt, \,
    \OsheafXt( -2 \tau_1 - \tau_2 + \phi) 
  \Big) = 0
  \,.
\end{equation}

\subsubsection{Cohomology of an Extension}

We can now compute the cohomology of the $SU(2)$ bundle $\Hsheaf$. By
definition, it is the middle term of a short exact sequence
\begin{equation}
  0 
  \longrightarrow
  \OsheafXt( 2 \tau_1 + \tau_2 - \phi)
  \longrightarrow
  \Hsheaf
  \longrightarrow
  \OsheafXt(-2\tau_1 - \tau_2 + \phi)  
  \longrightarrow
  0
  \,.
\end{equation}
The associated long exact sequence of cohomology groups reads
\begin{equation}
  \cdots
  \underbrace{
    H^i\big(\Xt,\,\OsheafXt( 2 \tau_1 + \tau_2 - \phi) \big)
  }_{=0}
  \to
  H^i\big( \Xt,\,\Hsheaf\big)
  \to
  \underbrace{
    H^i\big(\Xt,\,\OsheafXt(-2\tau_1 - \tau_2 + \phi) \big)  
  }_{=0}
  \cdots
  \,,
\end{equation}
so we immediately obtain
\begin{equation}
  \label{eq:Hcohvanish}
  H^\ast\big(\Xt,\,\Hsheaf\big) = 0 
  \,.
\end{equation}

\subsection{Absence of Hidden Matter}
\label{sec:Hmatter}

Let us return to the physical application of the $SU(2)$ bundle
$\Hsheaf$ which we just constructed. We are going to take the usual
regular embedding of $SU(2)$ in $E_8$ with commutant $E_7$. The fiber
product of the $SU(2)$ principal bundle together with the trivial
$E_7$ principal bundle then determines an $E_8$ principal bundle,
which we take to be our hidden $E_8$ gauge bundle.

Now, the gauge fermions in the heterotic string transform in the
adjoint representation of $E_8$, which branches as
\begin{equation}
  R[E_8] \owns~
  \Rep{248}
  =
  (\Rep{3}, \Rep{1}) \oplus
  (\Rep{1}, \Rep{133}) \oplus 
  (\Rep{2}, \Rep{56}) 
  ~\in R\big[ SU(2) \times E_7 \big]
\end{equation}
Correspondingly, the fermions, that is, the rank $248$ vector bundle
$\Esheaf_8^H$ associated to the hidden $E_8$ principal bundle,
decompose as 
\begin{equation}
  \Esheaf_8^H = 
  \Big( \Sym^2(\Hsheaf) \otimes \theta(1) \Big)
  \oplus 
  \Big( \theta(1) \otimes \theta(133) \Big)
  \oplus 
  \Big( \Hsheaf \otimes \theta(56) \Big)
  \,,
\end{equation}
where $\theta(n)\eqdef \Xt\times \C^n$ denotes the rank $n$ trivial
vector bundle.

\subsection{The Space of Extensions}
\label{sec:Hext}

It remains to determine the space of extensions for the short exact
sequence eq.~\eqref{eq:Hdef}. Using elementary properties of the
global Ext, this is given by
\begin{multline}
  \Ext^1\Big( 
  \OsheafXt(-2\tau_1 - \tau_2 + \phi)  
  ,~
  \OsheafXt( 2 \tau_1 + \tau_2 - \phi)
  \Big) 
  = \\ =
  \Ext^1\Big( 
  \OsheafXt
  ,~
  \OsheafXt( 4 \tau_1 + 2 \tau_2 - 2\phi)
  \Big) = 
  H^1\Big( 
  \Xt
  ,~
  \OsheafXt( 4 \tau_1 + 2 \tau_2 - 2\phi)
  \Big) 
  \,.
\end{multline}
From a \LSS{}, we can easily compute the cohomology groups of this
line bundle and obtain
\begin{equation}
  \label{eq:HOXt4t12t22pdimension}
\begin{gathered}
  \dim_\C H^p\Big(\Xt,\, 
  \OsheafXt( 4 \tau_1 + 2 \tau_2 - 2\phi)
  \Big) = 
  \begin{cases}
    54 & p=1 \\
    0  & p\not= 1 
    \,.
  \end{cases}
\end{gathered}
\end{equation}
We could now trace the $\ZZZ$ action through the \LSS{} and determine
directly which $54$-dimensional representation occurs. However, there
is a simple shortcut which we will employ instead.

For this, note that the index
\begin{equation}
\begin{split}
  &
  \idx
  \Big( 
  \OsheafXt( 4 \tau_1 + 2 \tau_2 - 2\phi)
  \Big)
  =
  \\ &\qquad=
  \sum_{p=0}^3
  (-1)^p
  \dim_\C H^p\Big(\Xt,\, 
  \OsheafXt( 4 \tau_1 + 2 \tau_2 - 2\phi)
  \Big) 
  = - 54      
\end{split}
\end{equation}
is simply divided by the order of the group $G=\ZZZ$ when descending
to the quotient. That is,
\begin{equation}
\begin{split}
  &
  \idx\Big( 
  \OsheafXt( 4 \tau_1 + 2 \tau_2 - 2\phi) \Big/ G
  \Big)
  =
  \\ &\qquad=
  \sum_{p=0}^3
  (-1)^p
  \dim_\C H^p\Big(\Xt,\, 
  \OsheafXt( 4 \tau_1 + 2 \tau_2 - 2\phi)
  \Big)^G
  = - 9
\end{split}
\end{equation}
Since from eq.~\eqref{eq:HOXt4t12t22pdimension} all the other
cohomology groups vanish, only the $p=1$ term can have an invariant
part which, moreover, must be $9$-dimensional.

The same reasoning can be applied to the line bundles $\chi \OsheafXt(
4 \tau_1 + 2 \tau_2 - 2\phi)$ for any character $\chi$ of $G$. The $G$
invariant subspace must always be $9$-dimensional. Of course, the
dimension of this invariant subspace is nothing else but the
multiplicity of the representation $\chi^{-1}$ in $H^1\big(\Xt,
\OsheafXt( 4 \tau_1 + 2 \tau_2 - 2\phi) \big)$. It follows that this
cohomology group decomposes as the sum of all nine irreducible $\ZZZ$
representations, each with multiplicity $6$:
\begin{equation}
  \label{eq:O4t12t2m2fcoh}
  H^p\Big(\Xt,\, \OsheafXt(4\tau_1+2\tau_2-2\phi) \Big) =
  \begin{cases}
    0
    \,, & p=3 \\
    0
    \,, & p=2 \\
    6 \Reg(\ZZZ)
    \,, & p=1 \\    
    0
    \,, & p=0\,. \\
  \end{cases}
\end{equation}
Here we used the fact that the \textdef{regular} representation of
$G=\ZZZ$, that is the representation of $G$ on its group ring $\C[G]$,
decomposes as
\begin{equation}
  \Reg(\ZZZ) = 
  \bigoplus_{i,j=0}^2 
  \chi_1^i \chi_2^j
  \,,
\end{equation}
the sum of every irreducible $G$ representation with multiplicity one.

To summarize, the $\Ext$ group in question is
\begin{equation}
  \Ext^1\Big( 
  \OsheafXt(-2\tau_1 - \tau_2 + \phi)  
  ,~
  \OsheafXt( 2 \tau_1 + \tau_2 - \phi)
  \Big) 
  = 6 \Reg(\ZZZ)
  \,.
\end{equation}

\subsection{Checks on Stability}
\label{sec:stabilityH}

As we stated in the introduction, in this paper we are not going to
prove stability of the vector bundles in a mathematically rigorous
sense. We will, however, subject them to the following two important
and nontrivial tests.

The first is that a stable vector bundle is necessarily
simple~\cite{HuybrechtsLehn}, that is, has no endomorphisms except
multiplication by a constant. This can be expressed as
\begin{equation}
  \label{eq:HEnd}
  \End(\Hsheaf) \simeq \C
  \quad \Leftrightarrow \quad
  H^0\Big(\Xt, \Hsheaf \otimes \Hsheafdual\Big)
  = 1
  \,.
\end{equation}
The other test for stability is 
\begin{equation}
  \label{eq:HH3zero}
  \text{\Hsheaf~stable} 
  \quad \Rightarrow \quad
  H^0\Big(\Xt,\Hsheafdual\Big)=0
  \,,
\end{equation}
which follows from the following contradiction. Assume that
$H^0(\Xt,\Hsheafdual)\not=0$. This means that there is a global
section of $\Hsheafdual$. But a global section of \Hsheafdual{} is a
map $s:\Hsheaf\to\Osheaf$, which is necessarily surjective and can be
completed to a short exact sequence
\begin{equation}
  \label{eq:H0Hsheafdualses}
  0 
  \longrightarrow
  \ker(s)
  \longrightarrow
  \Hsheaf
  \stackrel{s}{\longrightarrow}
  \Osheaf
  \longrightarrow
  0
  \,.
\end{equation}
Since $\Hsheaf$ has vanishing first Chern class, the slopes $\mu$ of the
bundles all vanish,
\begin{equation}
  \mu\big( \ker(s) \big) = 
  \frac{
    c_1\big( \ker(s) \big) 
  }{
    \rank\big( \ker(s) \big) 
  }
  = 0 = \mu\big( \Hsheaf \big)
  \,.
\end{equation}
Therefore $\ker(s)$ would be a destabilizing subsheaf of $\Hsheaf$,
and $\Hsheaf$ could at most be semistable.

Finally, note that the dual bundle of a stable bundle is a again
stable. Therefore, (slope-)stability of $\Hsheaf$ implies the
following three constraints on cohomology groups:
\begin{equation}
  \label{eq:H:check}
  H^0\Big(\Xt,\, \Hsheaf \otimes \Hsheafdual \Big) = 1
  \,,\qquad
  H^0\Big(\Xt,\, \Hsheaf \Big) = 0
  \,,\qquad
  H^0\Big(\Xt,\, \Hsheaf^\dual \Big) = 0
  \,.
\end{equation}
We already computed the cohomology of $\Hsheaf$ (and $\Hsheaf^\dual$
by Serre duality) and found that it vanishes, see
eq.~\eqref{eq:Hcohvanish}. It remains to compute the cohomology of
$\Hsheaf \otimes \Hsheafdual$, which we will do in the following
Subsection. We find that, indeed, the vector bundle $\Hsheaf$ is
simple, that is, $H^0\big(\Xt,\, \Hsheaf \otimes \Hsheafdual \big) =
1$. Hence, $\Hsheaf$ passes all the checks on stability.

\subsection{Simplicity}
\label{sec:simple}

Let us perform the promised computation of the endomorphisms of the
extension eq.~\eqref{eq:Hdef}. We will first discuss the general case
of an arbitrary extension of vector bundles
\begin{equation}
  \label{eq:ABCext}
  0 
  \longrightarrow
  \Asheaf 
  \longrightarrow
  \Bsheaf 
  \longrightarrow
  \Csheaf 
  \longrightarrow
  0
  \,.
\end{equation}
Then the dual vector bundle $\Bsheaf^\dual$ fits into the short exact
sequence
\begin{equation}
  0 
  \longrightarrow
  \Csheaf^\dual
  \longrightarrow
  \Bsheaf^\dual 
  \longrightarrow
  \Asheaf^\dual
  \longrightarrow
  0
\end{equation}
and the tensor product $\Bsheaf \otimes \Bsheaf^\dual$ fits into the
commutative diagram
\begin{equation}
  \label{eq:ABC3x3diag}
  \vcenter{\xymatrix{
      &
      0 \ar[d] &
      0 \ar[d] &
      0 \ar[d] &
      \\
      0 \ar[r] &
      \Asheaf \otimes \Csheaf^\dual \ar[d] \ar[r] &
      \Asheaf \otimes \Bsheaf^\dual \ar[d] \ar[r] &
      \Asheaf \otimes \Asheaf^\dual \ar[d] \ar[r] &
      0
      \\
      0 \ar[r] &
      \Bsheaf \otimes \Csheaf^\dual \ar[d] \ar[r] &
      \Bsheaf \otimes \Bsheaf^\dual \ar[d] \ar[r] &
      \Bsheaf \otimes \Asheaf^\dual \ar[d] \ar[r] &
      0
      \\
      0 \ar[r] &
      \Csheaf \otimes \Csheaf^\dual \ar[d] \ar[r] &
      \Csheaf \otimes \Bsheaf^\dual \ar[d] \ar[r] &
      \Csheaf \otimes \Asheaf^\dual \ar[d] \ar[r] &
      0
      \\
      &
      0 &
      0 &
      0 &
    }}
\end{equation}
with all rows and columns exact. These nested short exact sequences
yield interrelated long exact sequences of cohomology groups which we
are going to use.

Let us apply these general considerations to the extension
eq.~\eqref{eq:Hdef}. The tensor products of the line bundles are
either the trivial line bundle or the line bundle
$\OsheafXt(4\tau_1+2\tau_2-2\phi)$, whose cohomology is noted in
eq.~\eqref{eq:O4t12t2m2fcoh}. Using these values, the commutative
diagram of long exact sequences simplifies to
\begin{equation}
  \label{eq:H3x3diag}
  \vcenter{\xymatrix@M=2mm@=7mm{
      &
      0 \ar[d] &
      0 \ar[d] &
      0 \ar[d] 
      \\
      0 \ar[r] &
      0 \ar[d] \ar[r] &
      *+<1 cm>[o][F]{} \ar[d] \ar[r] &
      1 \ar[d] \ar[r]^<>(0.5){\delta} &
      6 \Reg(\ZZZ) \ar[r] &
      \cdots
      \\
      0 \ar[r] &
      *+<1 cm>[o][F]{} \ar[d] \ar[r] &
      H^0\Big( \Xt,\, \Hsheaf \otimes \Hsheaf^\dual \Big) \ar[d] \ar[r] &
      *+<1 cm>[o][F]{} \ar[d] \ar[r] &     
      \cdots \hspace{15mm}
      \\
      0 \ar[r] &
      1 \ar[d]_{\delta} \ar[r] &
      *+<1 cm>[o][F]{} \ar[d] \ar[r] &
      0 \ar[d] \ar[r] &
      \cdots \hspace{15mm}
      \\
      &
      *+<2mm>{\strut} \save[]*{6 \Reg(\ZZZ)} \restore \ar[d] &
      \vdots &
      \vdots &
      \\
      & 
      \vdots
    }}
\end{equation}
where the empty circles are cohomology groups (of tensor products of
$\Hsheaf$ and line bundles) that are yet to be determined. To do this,
we must analyze the coboundary map
\begin{equation}
  \delta: 
  H^0\Big( \Xt,\, \OsheafXt \Big) \to 
  H^1\Big( \Xt,\, \OsheafXt(4\tau_1+2\tau_2-2\phi) \Big) 
  \,.
\end{equation}
This is simply multiplication by the extension
class\footnote{The discussion of the extension class is unavoidable at
  this point. If the extension were trivial, that is, $\epsilon=0$,
  then $\Hsheaf$ would be a sum of two line bundles. But such a sum is
  never simple as the two line bundles can be scaled independently.}
\begin{equation}
\begin{split}
  \epsilon \in 
  &\, 
  \Ext^1\Big( 
  \OsheafXt(-2\tau_1 - \tau_2 + \phi)  
  ,\,
  \OsheafXt( 2 \tau_1 + \tau_2 - \phi)
  \Big) 
  = \\ & \hspace{4cm} =
  H^1\Big( 
  \Xt
  ,\,
  \OsheafXt( 4 \tau_1 + 2 \tau_2 - 2\phi)
  \Big)   
  \,,  
\end{split}
\end{equation}
that is, the cohomology class encoding the choice of extension in
eq.~\eqref{eq:Hdef}. By our assumption that the extension class is
generic (that is, not zero), the coboundary map $\delta$ is injective.
This determines the missing entries in the long exact sequences
eq.~\eqref{eq:H3x3diag} to be
\begin{equation}
  \label{eq:H3x3diagfinal}
  \vcenter{\xymatrix@M=2mm@=5mm{
      &
      0 \ar[d] &
      0 \ar[d] &
      0 \ar[d] 
      \\
      0 \ar[r] &
      0 \ar[d] \ar[r] &
      0 \ar[d] \ar[r] &
      1 \ar[d] \ar[r]^<>(0.5){\delta} &
      \cdots
      \\
      0 \ar[r] &
      0 \ar[d] \ar[r] &
      H^0\Big( \Xt,\, \Hsheaf \otimes \Hsheaf^\dual \Big) \ar[d] \ar[r] &
      1 \ar[d] \ar[r] &     
      \cdots
      \\
      0 \ar[r] &
      1 \ar[d]_<>(0.5){\delta} \ar[r] &
      1 \ar[d] \ar[r] &
      0 \ar[d] \ar[r] &
      \cdots
      \\
      &
      \vdots &
      \vdots &
      \vdots
    }}
\end{equation}
Exactness then implies that the desired $H^0$ is either zero or one
dimensional. But there is always a global section of $\Hsheaf \otimes
\Hsheaf^\dual$ corresponding to multiplication by an overall constant.
Hence,
\begin{equation}
  H^0\Big( \Xt,\, \Hsheaf \otimes \Hsheaf^\dual \Big)
  = 1
  \,.
\end{equation}

\section{The Visible \texorpdfstring{$\mathbf{E_8}$}{E8} Bundle}
\label{sec:visibeE8}

\subsection{The \texorpdfstring{$SU(4)$}{SU(4)} Bundle}
\label{sec:SU4bundle}

Let us now construct the $SU(4)$ instanton inside the visible $E_8$
gauge group. We first define an auxiliary bundle $\Wsheaf$ on the
$dP_9$ surface $B_2$. For that, we take an extension of the form
\begin{equation}
  \label{eq:Wbundle}
  0 
  \longrightarrow
  \OsheafBtwo(-2 f) 
  \longrightarrow
  \Wsheaf
  \longrightarrow
  \chi_2
  \OsheafBtwo(2 f) \otimes I_9
  \longrightarrow
  0
  \,,
\end{equation}
where $I_9$ denotes the ideal sheaf of $9$ points, which we take to be
one generic $G$ orbit. Then the $9$ points end up in $3$ distinct
fibers of the elliptic fibration $\beta_2:B_2\to \CP^1$, each
containing $3$ points. Furthermore, we remark that the first Chern
class of $\Wsheaf$ vanishes, and hence $\Wsheaf\simeq \Wsheaf^\vee$. 

Using the pullback of $\Wsheaf$, we now define two $U(2)$ bundles on
$\Xt$ as 
\begin{equation}
  \label{eq:V12bundles}
  \begin{split}
    \Vsheaf_1 ~&\eqdef~
    \chi_2 \OsheafXt(-\tau_1+\tau_2) \oplus 
    \chi_2 \OsheafXt(-\tau_1+\tau_2) =   
    2 \chi_2 \OsheafXt(-\tau_1+\tau_2)
    \,,
    \\
    \Vsheaf_2 ~&\eqdef~
    \OsheafXt(\tau_1-\tau_2) \otimes \pi_2^\ast(\Wsheaf)
    \,.
  \end{split}
\end{equation}
Finally, we define the rank $4$ bundle $\Vsheaf$ as a generic
extension of $\Vsheaf_2$ with $\Vsheaf_1$, that is,
\begin{equation}
  \label{eq:Vbundle}
  0 
  \longrightarrow
  \Vsheaf_1
  \longrightarrow
  \Vsheaf
  \longrightarrow
  \Vsheaf_2
  \longrightarrow
  0
  \,.
\end{equation}
Because the first Chern classes of the $U(2)$ bundles add up to zero,
$c_1(\Vsheaf_1)+c_1(\Vsheaf_2)=0=c_1(\Vsheaf)$, the bundle $\Vsheaf$
defines an $SU(4)$ gauge bundle.

Simply computing the Chern character, see eq.~\eqref{eq:chvisible}, we
immediately can conclude that the net number of generations (the
index) on the covering space is
\begin{equation}
\begin{split}
  N_{\text{gen}}\big(\Xt\big) =&\, 
  \idx(\Vsheaf) =
  \int_\Xt 
  ch\big(\Vsheaf\big)\, Td\big(T\Xt\big)
  = \\ =&\,
  - \int_\Xt 9 \, PD\big(\tau_1\tau_2^2\big) = 
  - \int_\Xt 9 \, PD\big(3 \ptset\big) = -27
  \,,
\end{split}
\end{equation}
where $PD$ denotes the Poincar\'e dual. Therefore, the net number of
generations on the quotient is
\begin{equation}
  N_\text{gen}\big(X\big) = 
  N_\text{gen}\big(\Xt/G\big) = 
  \frac{1}{|G|} 
  N_\text{gen}\big(\Xt\big) = -3
  \,,
\end{equation}
and we do get $3$ net generations (the sign of the index is
irrelevant). Of course, obtaining $3$ net generations is necessary but
not sufficient to get a standard model spectrum. In general, one can
expect a whole zoo of exotic matter accompanying these $3$
generations. To discuss these, we must compute the actual cohomology
groups which correspond to the massless modes of the Dirac operator,
and not just their alternating sum. We will compute the cohomology
groups of $\Vsheaf$ and $\wedge^2\Vsheaf$ in the remainder of this
section, and then extract the complete low energy spectrum in
Section~\ref{sec:spectrum}.

\subsection{Anomaly cancellation with Five-Branes}
\label{sec:withfivebranes}

First of all, let us check the heterotic anomaly cancellation. It
requires that 
\begin{equation}
  \big[ \tr R^2 \big]
  - 
  \frac{1}{30} \big[ \tr F^2 \big]
  = 0 ~\in H^4\big(\Xt,\, \Z\big)
  \,,
\end{equation}
where $F$ is in the adjoint representation of the $E_8\times E_8$
gauge group. For any regular $SU(n)$ subgroup of $E_8$, the second
Chern class of an $SU(n)$ vector bundle $\Fsheaf$ and the second Chern
class of the associated adjoint $E_8$ bundle $\Esheaf_8$ are related
by
\begin{equation}
  c_2\big( \Fsheaf \big) =
  \frac{1}{60} c_2\big( \Esheaf_8 \big)   
  \,.
\end{equation}
Now, we pick a $SU(n_V)$ bundle $\Vsheaf$ and a $SU(n_H)$ bundle
$\Hsheaf$ and construct an $E_8^V\times E_8^H$ bundle from the regular
embeddings of $SU(n_i)\subset E_8^i$. Then we can rewrite the anomaly
cancellation in terms of characteristic classes as 
\begin{equation}
  c_2\big(T\Xt\big) 
  - \frac{1}{60} c_2\big(\Esheaf_8^V\big) 
  - \frac{1}{60} c_2\big(\Esheaf_8^H\big)
  =
  c_2\big(T\Xt\big) 
  - c_2\big(\Vsheaf\big) - c_2\big(\Hsheaf\big)
  = 0
  ~\in H^4\big(\Xt,\Z\big)
  \,.
\end{equation}
This condition can be slightly relaxed if one allows for five-branes,
which also contribute to the anomaly. To preserve supersymmetry, the
five-branes must be wrapped on an effective curve, that is an actual
holomorphic curve rather than a sum of curves and orientation-reversed
curves. The Poincar\'e dual of the curve $C$ then contributes to the
anomaly as
\begin{equation}
  \label{eq:HetAnomaly}
  c_2\big(T\Xt\big) 
  - c_2\big(\Vsheaf\big) - c_2\big(\Hsheaf\big)
  = PD(C)
  \in H^4\big(\Xt,\Z\big)
  \,.
\end{equation}
If this equation holds, then wrapping a five-brane on the curve $C$
cancels the heterotic anomaly. Now, the Chern classes of (the tangent
bundle of) a fiber product of $dP_9$ surfaces was already computed
in~\cite{dP9torusfib}. One finds that
\begin{equation}
  c_1\big(T\Xt\big) = 0
  \,, \quad
  c_2\big(T\Xt\big) = 12\left( \tau_1^2 + \tau_2^2 \right)
  \,, \quad
  c_3\big(T\Xt\big) = 0
  \,.
\end{equation}
The Chern classes for the visible and the hidden gauge bundle are also
easy to compute. For simplicity, we work with the Chern
character. The Chern character of the hidden bundle is 
\begin{equation}
\label{eq:chhidden}
\begin{split}
  ch\big(\Hsheaf\big) &= 
  ch\Big( \Osheaf( 2 \tau_1 + \tau_2 - \phi) \Big) +
  ch\Big( \Osheaf(-2 \tau_1 - \tau_2 + \phi) \Big)
  = 
  \\ &=
  e^{2 \tau_1 + \tau_2 - \phi} + 
  e^{-2 \tau_1 - \tau_2 + \phi} 
  =
  2 - 8\tau_1^2 - 5\tau_2^2 + 4\tau_1\tau_2
  \,,
\end{split}
\end{equation}
where we have used the relations in the intersection ring,
eq.~\eqref{eq:intersectionRing}. The second Chern class is then
\begin{equation}
  \label{eq:c2hidden}
  c_2(\Hsheaf) = \frac{1}{2} c_1(\Hsheaf)^2 - ch_2(\Hsheaf) 
  = 8\tau_1^2 + 5\tau_2^2 - 4\tau_1\tau_2
  \,.
\end{equation}
Similarly, the Chern character of the visible bundle is given by
\begin{equation}
\label{eq:chvisible}
\begin{split}
  ch\big(\Vsheaf\big) &= 
  ch\big(\Vsheaf_1\big) + 
  ch\big(\Vsheaf_2\big) 
  =
  2 ch\Big( \Osheaf( - \tau_1 + \tau_2 ) \Big) +
  ch\Big( \Osheaf(\tau_1-\tau_2) \Big) ch\big(\pi_2^\ast W\big)
  = \\ &=
  2 e^{-\tau_1+\tau_2} + 
  e^{+\tau_1-\tau_2} 
  \Big( e^{-2\phi} + e^{2\phi} \big(1-9\tau_2^2) \Big)
  = \\ &=
  4 + 2 \tau_1^2 -7\tau_2^2 - 4\tau_1\tau_2 -9 \tau_1\tau_2^2
\end{split}
\end{equation}
and its second Chern class is 
\begin{equation}
  \label{eq:c2visible}
  c_2\big(\Vsheaf\big) = 
  - 2 \tau_1^2 +7\tau_2^2 + 4\tau_1\tau_2
  \,.
\end{equation}
Combining everything, the combined gravity and gauge contribution to
the anomaly is
\begin{equation}
  \label{eq:TXVHanomaly}
  c_2\big( T\Xt \big) 
  - c_2\big( \Vsheaf \big) 
  - c_2\big( \Hsheaf \big)
  = 6 \tau_1^2
  \,.
\end{equation}
The homology class 
\begin{equation}
  PD(\tau_1^2)=
  PD\big(\pi_1^{-1}(t^2)\big)=
  PD\Big( \pi_1^{-1}\big(\ptset\big) \Big)
\end{equation}
is effective, indeed it is a multiple of an elliptic fiber of
$\pi_1$. Wrapping five-branes on this homology class cancels the
anomaly, eq.~\eqref{eq:TXVHanomaly}, and yields a completely well
defined, albeit strongly coupled, compactification.
  
So far, we really worked on the universal covering space $\Xt$,
whereas we ultimately want to compactify on the $G\simeq\ZZZ$ quotient
$X=\Xt/G$. This is justified as follows. By definition, we have 
a quotient map
\begin{equation}
  q: \Xt \to X
  ,~
  p \mapsto Gp
  \,.
\end{equation}
Moreover, the vector bundle $\Vsheaf$ on $\Xt$ is really the pull back
of a vector bundle $\Vsheaf/G$ on $X$. The Chern classes are natural,
that is
\begin{equation}
  \vcenter{\xymatrix{
      \Vsheaf \ar[d]  & 
      \Vsheaf/G \ar[d] \ar[l]^{q^\ast} \\
      \Xt \ar[r]^q &
      X
    }}
  \quad
  \Rightarrow 
  \quad
  c_i\big( \Vsheaf \big) = 
  c_i\Big( q^\ast \big(\Vsheaf/G \big) \Big) = 
  q^\ast c_i\Big( \Vsheaf/G \Big)
  \,.
\end{equation}
At least rationally\footnote{We are ignoring possible torsion issues
  for the purposes of this paper.}, the pullback of cohomology classes
\begin{equation}
  q^\ast: H^{2i}\Big( X,\, \Q\Big) \to 
  H^{2i}\Big( \Xt, \Q\Big)
\end{equation}
is just the inclusion of the $G$ invariant cohomology,
\begin{equation}
  H^{2i}\Big( X,\, \Q\Big) =
  H^{2i}\Big( \Xt,\, \Q\Big)^G
  \subset 
  H^{2i}\Big( \Xt, \Q\Big)
  \,.
\end{equation}
In particular, $q^\ast$ is injective. Hence, the vanishing of a sum of
Chern classes on $\Xt$ is sufficient to conclude that the same sum on
the quotient also vanishes.

\subsection{Cohomology of \texorpdfstring{$\Vsheaf$}{V}}
\label{sec:cohomologyV}

\subsubsection{Cohomology of \texorpdfstring{$\Vsheaf_1$}{V1}}

First, let us consider the cohomology of the line bundle $\chi_2
\Osheaf(-\tau_1+\tau_2)$, one half of the vector bundle $\Vsheaf_1$.
We compute the cohomology by pushing down to $B_1$ and obtain for the
$E_2$ tableau of the \LSS
\begin{equation}
\begin{split}
  E_2^{p,q}\big( \Xt|B_1 \big) 
  =&\,
  H^p\Big( B_1,\,
    R^q\pi_{1\ast} 
    \big(
      \chi_2
      \Osheaf(-\tau_1+\tau_2) 
    \big)
  \Big)
  = \\ =&\,
  H^p\Big( B_1,\,
    \OsheafBone(-t) \otimes R^q\pi_{1\ast} 
    \pi_2^\ast
    \big(
      \chi_2
      \OsheafBtwo(t)
    \big)
  \Big)
  = \\ =&\,
  H^p\Big( B_1,\,
    \OsheafBone(-t) \otimes \beta_1^\ast R^q\beta_{2\ast} 
    \big(
      \chi_2
      \OsheafBtwo(t)
    \big)
  \Big)
  \\ \Rightarrow& \quad
  H^{p+q}\Big( \Xt,\, 
    \chi_2
    \Osheaf\big(-\tau_1+\tau_2\big)
  \Big)
  \,.
\end{split}
\end{equation}
Because the fiber degree of $\OsheafBtwo(t)$ is positive, the $q=1$
row vanishes, while for $q=0$ we obtain
\begin{equation}
\begin{split}
  H^p\Big( B_1,\,
    \OsheafBone(-t) \otimes \beta_1^\ast \beta_{2\ast} 
    \big(
      \chi_2
      \OsheafBtwo(t)
    \big)
  \Big)
  =&\,
  H^p\Big( B_1,\,
    \OsheafBone(-t) \otimes \beta_1^\ast 
    \big(
      3 \chi_2
      \OsheafP
    \big)
  \Big)
  = \\ =&\,
  H^p\Big( B_1,\,
  3 \chi_2
  \OsheafBone(-t)
  \Big)
  \,.
\end{split}
\end{equation}
We compute this cohomology group by another \LSS{}
\begin{equation}
\begin{split}
  E_2^{p,q}\big(B_1|\CP^1\big)
  = &\,
  H^p\Big( \CP^1,\,
  3 R^q \beta_{1\ast} 
  \big(
  \chi_2
  \OsheafBone(-t)
  \big)
  \Big) 
  = \\ =&\,
  \begin{cases}
    H^p\Big( \CP^1,\,
    9 \chi_1 \chi_2 \OsheafP(-1)
    \Big)     
    \,, & q=1
    \\
    0
    \,, & q=0
    \,.
  \end{cases}
\end{split}
\end{equation}
Every entry in the second tableau vanishes and, therefore, all
cohomology groups vanish,
\begin{equation}
  H^\ast\Big( \Xt,\, \Vsheaf_1 \Big) = 0
  \,.
\end{equation}

\subsubsection{Cohomology of \texorpdfstring{$\Vsheaf_2$}{V2}}

We continue with the cohomology of the rank $2$ bundle $\Vsheaf_2$.
Its dual\footnote{\label{note:Wdual}We remark that
  $\Rep{2}=\barRep{2}\in R[SU(2)]$ and, therefore,
  $\Wsheaf\simeq\Wsheaf^\dual$ as vector bundles. However, the dual
  equivariant structure differs, and $\Wsheaf^\dual$ is an extension
  \begin{equation}
    \label{eq:Wdualbundle}
    0 
    \longrightarrow
    \chi_2^2
    \OsheafBtwo(-2 f) 
    \longrightarrow
    \Wsheaf^\dual
    \longrightarrow
    \OsheafBtwo(2 f) \otimes I_9
    \longrightarrow
    0
    \,.
  \end{equation}
} is
\begin{equation}
  \label{eq:V2dualbundle}
  \Vsheaf_2^\dual =
  \Osheaf(-\tau_1+\tau_2) \otimes \pi_2^\ast(\Wsheaf^\dual)
  \,,
\end{equation}
and, because the degree on the fiber is negative, the pushdowns
\begin{equation}
  \pi_{1\ast} \big( \Vsheaf_2 \big) = 0  
  \,, \qquad 
  \pi_{2\ast} \big( \Vsheaf_2^\dual \big) = 0  
\end{equation}
automatically vanish. But a global sections on $\Xt$ is a choice of
section on every fiber, that is,
\begin{equation}
  H^0\Big(\Xt,\, \Vsheaf_i\Big) = 
  H^0\Big(B_j,\, \pi_{j\ast}\big(\Vsheaf_i\big) \Big)
  \,,\qquad
  i=1,2
  \,,~
  j=1,2
  \,,
\end{equation}
and we immediately conclude that
\begin{equation}
  H^0\Big(\Xt,\, \Vsheaf_2 \Big) = 0
  \,,\qquad
  H^3\Big(\Xt,\, \Vsheaf_2 \Big) \simeq
  H^0\Big(\Xt,\, \Vsheaf_2^\dual \Big)^\dual = 0
  \,.
\end{equation}
Next, let us compute $H^2\big(\Xt, \Vsheaf_2\big)$ or, rather, its
Serre dual $H^1\big(\Xt, \Vsheaf_2^\dual\big)$. Using the \LSS{}, one
immediately shows that
\begin{equation}
  H^1\Big(\Xt,\, \Vsheaf_2^\dual\Big) =
  H^0\Big(B_2,\, R^1\pi_{2\ast} \Vsheaf_2^\dual \Big) =
  H^0\Big(\CP^1,\, \beta_{2\ast} 
  \big( R^1\pi_{2\ast} \Vsheaf_2^\dual\big) \Big)
  \,.
\end{equation}
We have to determine the higher pushdown of $\Vsheaf_2^\dual$. For
that, we start with the projection formula and arrive at
\begin{equation}
  \label{eq:V2pushdown}
\begin{split}
  \beta_{2\ast} \big( R^1\pi_{2\ast} \Vsheaf_2^\dual\big)
  =&\,
  \beta_{2\ast} \Big( 
  R^1\pi_{2\ast} \big( \OsheafXt(-\tau_1) \big) 
  \otimes \OsheafBtwo(t) \otimes \Wsheaf^\dual
  \Big)  
  =\\=&\,
  \beta_{2\ast} \Big( 
  R^1\pi_{2\ast} \circ \pi_1^\ast \big( \OsheafBone(-t) \big) 
  \otimes \OsheafBtwo(t) \otimes \Wsheaf^\dual
  \Big)  
  =\\=&\,
  \beta_{2\ast} \Big( 
  \beta_2^\ast \circ R^1\beta_{1\ast} \big( \OsheafBone(-t) \big) 
  \otimes \OsheafBtwo(t) \otimes \Wsheaf^\dual
  \Big)    
  =\\=&\,
  R^1\beta_{1\ast} \Big( \OsheafBone(-t) \Big) \otimes
  \beta_{2\ast} \Big(
  \OsheafBtwo(t) \otimes \Wsheaf^\dual
  \Big)    
  \,.
\end{split}
\end{equation}
Since the pushdown does not distribute over tensor products, we have
to compute the second factor separately. For that, we use the long
exact sequence associated to the pushdown of the short exact sequence
eq.~\eqref{eq:Wbundle}. But that sequence contains the ideal sheaf,
whose pushdown we have to determine first. By definition, the ideal
sheaf is the sheaf of functions vanishing at the given $9$ points,
that is, the kernel of the restriction to the skyscraper sheaf on these
$9$ points,
\begin{equation}
  \label{eq:I9ses}
  0
  \longrightarrow
  I_9
  \longrightarrow
  \OsheafBtwo
  \longrightarrow
  \bigoplus_{i=1}^{9} \Osheaf_{p_i}
  \longrightarrow
  0
  \,.  
\end{equation}
Since we ultimately want to compute the pushdown of $\OsheafBtwo(t)
\otimes \Wsheaf$ instead of $\Wsheaf$, we tensor everything with
$\OsheafBtwo(t)$ and obtain
\begin{equation}
  \label{eq:OtI9ses}
  0
  \longrightarrow
  I_9 \otimes \OsheafBtwo(t)
  \longrightarrow
  \OsheafBtwo (t)
  \longrightarrow
  \bigoplus_{i=1}^{9} \Osheaf_{p_i}
  \longrightarrow
  0
  \,.  
\end{equation}
Now a generic $G$ orbit consists of $9$ points, living on $3$ distinct
fibers of the elliptic fibration $B_2$. The generator $g_1$, acting
nontrivially on the base, permutes the $3$ fibers. And the generator
$g_2$, acting by translation along each fiber separately, permutes the
triple of points on each fiber. We label the points such that
\begin{equation}
\begin{split}
  g_1: &~
  \vcenter{\xymatrix@!0{ p_1 \ar[rr] & & p_2 \ar[dl] \\ & p_3 \ar[ul] }}
  ,\quad
  \vcenter{\xymatrix@!0{ p_4 \ar[rr] & & p_5 \ar[dl] \\ & p_6 \ar[ul] }}
  ,\quad
  \vcenter{\xymatrix@!0{ p_7 \ar[rr] & & p_8 \ar[dl] \\ & p_9 \ar[ul] }}    
  \\
  g_2: &~
  \vcenter{\xymatrix@!0{ p_1 \ar[rr] & & p_4 \ar[dl] \\ & p_7 \ar[ul] }}
  ,\quad
  \vcenter{\xymatrix@!0{ p_2 \ar[rr] & & p_5 \ar[dl] \\ & p_8 \ar[ul] }}
  ,\quad
  \vcenter{\xymatrix@!0{ p_3 \ar[rr] & & p_6 \ar[dl] \\ & p_9 \ar[ul] }}
\end{split}
\end{equation}
The image under the projection map $\beta_2$ is then 
\begin{equation}
  \beta_2(p_{k}) = 
  \beta_2(p_{3+k}) = 
  \beta_2(p_{6+k})
  \,,\quad k=1,2,3
  \,,
\end{equation}
and accordingly
\begin{equation}
  \beta_{2\ast} \Big( \Osheaf_{p_{k}} \Big)
  =
  \beta_{2\ast} \Big( \Osheaf_{p_{3+k}} \Big)
  =
  \beta_{2\ast} \Big( \Osheaf_{p_{6+k}} \Big)
  \,,\quad k=1,2,3
  \,.  
\end{equation}
Furthermore, the fiber degree of $I_9 \otimes \OsheafBtwo(t)$ is
always positive or zero. This means that, for generic positions of the
points $p_k$, the first cohomology group of the restriction $\big( I_9
\otimes \OsheafBtwo(t)\big)|_f$ vanishes. Hence,
\begin{equation}
  R^1\beta_{2\ast} \Big( I_9 \otimes \OsheafBtwo(t) \Big) = 0
  \,.
\end{equation}
Using all of this, the long exact sequence for the pushdown of
eq.~\eqref{eq:OtI9ses} simplifies to 
\begin{equation}
  \label{eq:OtI9les}
  0
  \longrightarrow
  \beta_{2\ast}\big( I_9 \otimes \OsheafBtwo(t) \big)
  \longrightarrow
  3 \OsheafP 
  \longrightarrow
  \bigoplus_{i=1}^{3} \Osheaf_{3 \beta_2(p_i)}      
  \longrightarrow
  0
  \,.
\end{equation}
Part of the Heisenberg group action on $\beta_{2\ast}\OsheafBtwo(t) =
3 \OsheafP$ permutes the line bundles, this uniquely fixes the
multi degrees of the pushdown of $I_9 \otimes \OsheafBtwo(t)$ to be
\begin{equation}
  \beta_{2\ast}\Big( I_9 \otimes \OsheafBtwo(t) \Big)
  = 
  \bigoplus_{i=1}^{3} \OsheafP(-3)
  \,.
\end{equation}

Now we can compute the pushdown of $\Wsheaf^\dual \otimes
\OsheafBtwo(t)$, which fits into a short exact sequence\footnote{Here
  we are suppressing the $\ZZZ$ characters, since we have not properly
  defined what we mean by $\chi_2 \OsheafBtwo(t)$. Indeed, there is no
  $\ZZZ$ action on $\OsheafBtwo(t)$, only on the bundles $\Vsheaf_1$
  and $\Vsheaf_2$.  }
\begin{equation}
  \label{eq:OtWbundle}
  0 
  \longrightarrow
  \OsheafBtwo(-2 f + t) 
  \longrightarrow
  \Wsheaf^\dual \otimes \OsheafBtwo(t)
  \longrightarrow
  \OsheafBtwo(2 f ) \otimes I_9 \otimes \OsheafBtwo(t)
  \longrightarrow
  0
  \,.
\end{equation}
Since $R^1\beta_{2\ast}\OsheafBtwo(-2 f + t) =0$ for degree reasons,
the associated long exact sequence for the pushdown simplifies to
\begin{equation}
  0 
  \longrightarrow
  3 \OsheafP(-2 ) 
  \longrightarrow
  \beta_{2\ast} \Big( \Wsheaf^\dual \otimes \OsheafBtwo(t) \Big) 
  \longrightarrow
  3 
  \OsheafP( -1 )
  \longrightarrow
  0
  \,.  
\end{equation}
There is no nontrivial extension and, hence,
\begin{equation}
  \beta_{2\ast} \Big( \Wsheaf^\dual \otimes \OsheafBtwo(t) \Big) =
  3 \OsheafP(-1 ) 
  \oplus
  3 \OsheafP(-2 )   
  \,.
\end{equation}
Finally, we found the pushdown in eq.~\eqref{eq:V2pushdown} to be
\begin{equation}
\begin{split}
  \label{eq:V2pushdown2}
  \beta_{2\ast} \big( R^1\pi_{2\ast} \Vsheaf_2^\dual\big)
  =&\,
  R^1\beta_{1\ast} \big( \OsheafBone(-t) \big) \otimes
  \beta_{2\ast} \Big(
  \OsheafBtwo(t) \otimes \Wsheaf^\dual
  \Big)    
  =\\=&\,
  3 \OsheafP(-1) \otimes 
  \Big( 3 \OsheafP(-1 ) \oplus 3 \OsheafP(-2 ) \Big)
  =\\=&\,
  9 \OsheafP(-2 ) \oplus 9 \OsheafP(-3 )
  \,.
  \end{split}
\end{equation}
But negative degree line bundles do not have global sections, that is,
\begin{equation}
  H^1\Big(\Xt,\, \Vsheaf_2^\dual\Big) =
  H^0\Big(\CP^1,\, \beta_{2\ast} 
  \big( R^1\pi_{2\ast} \Vsheaf_2^\dual\big) \Big)
  = 
  0
  \,.
\end{equation}

\subsubsection{Cohomology of the Extension}

Thus far we determined that all cohomology groups of $\Vsheaf_1$ and
$\Vsheaf_2$, except $H^1(\Xt,\Vsheaf_2)$, vanish. We could compute
this last cohomology group again via the \LSS, using the formula in
eq.~\eqref{eq:V2pushdown2}. It is simpler to just use the index
\begin{equation}
  \idx(\Vsheaf) = \idx(\Vsheaf_1) + \idx(\Vsheaf_2) = -27
  \quad \Rightarrow \quad
  \idx(\Vsheaf_2) = -27
  \,.
\end{equation}
Together with the argument in Section~\ref{sec:Hext}, this determines
the dimension and the $\ZZZ$ representation to be
\begin{equation}
  \dim_\C
  H^1\big(\Xt,\,\Vsheaf_2\big)
  = 27
  \,,\qquad
  H^1\big(\Xt,\,\Vsheaf_2\big)
  = 
  3 \Reg(\ZZZ)
  \,.
\end{equation}

Now it is a simple matter to apply the long exact sequence associated
to the extension eq.~\eqref{eq:Vbundle} and find the cohomology groups
of $\Vsheaf$. Because so many entries vanish, there are no ambiguities
and we immediately obtain
\begin{equation}
  \label{eq:Vcoh}
  H^p\Big(\Xt,\, \Vsheaf \Big) =
  \begin{cases}
    0
    \,, & p=3 \\
    0
    \,, & p=2 \\
    3 \Reg(\ZZZ)
    \,, & p=1 \\    
    0
    \,, & p=0\,. \\
  \end{cases}
\end{equation}

\subsection{Cohomology of \texorpdfstring{$\wedge^2\Vsheaf$}{wedge2 V}}
\label{sec:cohomologyV2}

\subsubsection{Exact Sequences}

To compute the cohomology of $\wedge^2\Vsheaf$, we have to relate it
to cohomology groups involving $\Vsheaf_1$ and $\Vsheaf_2$. For this,
note that we are, by definition, given an injection
$\Vsheaf_1\to\Vsheaf$ and a surjection $\Vsheaf\to\Vsheaf_2$. From
these, we can construct maps of various tensor operations. In particular,
we get short exact sequences
\begin{subequations}
\begin{gather}
  0 
  \longrightarrow
  \wedge^2 \Vsheaf_1
  \longrightarrow
  \wedge^2 \Vsheaf
  \longrightarrow
  \Qsheaf_1
  \longrightarrow
  0
  \,, \\ 
  0 
  \longrightarrow
  \Qsheaf_2
  \longrightarrow
  \wedge^2 \Vsheaf
  \longrightarrow
  \wedge^2 \Vsheaf_2
  \longrightarrow
  0
\end{gather}
\end{subequations}
with some cokernel $\Qsheaf_1$ and kernel $\Qsheaf_2$. Now
$\wedge^2\Vsheaf$ contains contributions of $\wedge^2\Vsheaf_1$,
$\Vsheaf_1\otimes \Vsheaf_2$, and $\wedge^2\Vsheaf_2$ (but, of course,
is not a direct sum of these). Keeping this in mind, we can relate the
two short exact sequences in a commutative diagram
\begin{equation}
  \label{eq:wedge2Vdiagram}
  \vcenter{\xymatrix{
      & 
      & 
      0 \ar[d]
      & 
      0 \ar[d]
      &
      \\
      0 \ar[r] &
      \wedge^2 \Vsheaf_1 \ar[r] \ar@{=}[d] &
      \Qsheaf_1 \ar[r] \ar[d] &
      \Vsheaf_1 \otimes \Vsheaf_2 \ar[r] \ar[d] &
      0
      \\
      0 \ar[r] &
      \wedge^2 \Vsheaf_1 \ar[r] &
      \wedge^2 \Vsheaf \ar[r] \ar[d] &
      \Qsheaf_2 \ar[r] \ar[d] &
      0
      \\
      & &
      \wedge^2 \Vsheaf_2 \ar@{=}[r] \ar[d] &
      \wedge^2 \Vsheaf_2 \ar[d] 
      \\
      & & 0 & 0
    }}
\end{equation}
with exact rows and columns. 

Furthermore, we can easily determine the line bundles
\begin{subequations}
  \begin{align}
    \wedge^2 \Vsheaf_1 =&\, \chi_2^2 \OsheafXt(-2 \tau_1 + 2 \tau_2)
    \,,
    \\
    \wedge^2 \Vsheaf_2 =&\, \OsheafXt(2 \tau_1 - 2 \tau_2)
  \end{align}
\end{subequations}
simply by computing their first Chern class. As one can easily
calculate using the \LSS{}, it turns out that these line bundles have
no cohomology,
\begin{equation}
  H^\ast\Big( \Xt,\, \wedge^2 \Vsheaf_1 \Big) 
  = 0 =
  H^\ast\Big( \Xt,\, \wedge^2 \Vsheaf_2 \Big)  
  \,.
\end{equation}
Because of this lucky coincidence, the long exact sequences in
cohomology for the diagram eq.~\eqref{eq:wedge2Vdiagram} identify
\begin{equation}
  H^p\Big( \Xt,\, \wedge^2 \Vsheaf \Big) \simeq
  H^p\Big( \Xt,\, \Qsheaf_1 \Big) \simeq
  H^p\Big( \Xt,\, \Qsheaf_2 \Big) \simeq
  H^p\Big( \Xt,\, \Vsheaf_1 \otimes \Vsheaf_2 \Big)
  \,.
\end{equation}
Hence, we have simplified the computation of the cohomology groups of
$\wedge^2\Vsheaf$ to the cohomology of the vector bundle
\begin{equation}
  \Vsheaf_1 \otimes \Vsheaf_2 
  = 
  2 \chi_2 \pi_2^\ast\big( \Wsheaf \big)
  = 
  2 \pi_2^\ast\big( \chi_2 \Wsheaf \big)
  \,.
\end{equation}

\subsubsection{A Pushdown Formula}

To compute the cohomology of $\Vsheaf_1\otimes\Vsheaf_2$, we first
have to determine the pushdown of $\Wsheaf$ to the base $\CP^1$. 

We start by pushing down the ideal sheaf. The long exact sequence for
the pushdown of the sequence eq.~\eqref{eq:I9ses} is
\begin{equation}
  \label{eq:I9les}
  \vcenter{\xymatrix@R=10pt@M=4pt@H+=22pt{
      0 \ar[r] & 
      \beta_{2\ast}\big( I_9 \big)
      \ar[r] &
      \OsheafP 
      \ar[r]^<>(0.5){r} &
      \bigoplus_{k=1}^{3} \Osheaf_{3 \beta_2(p_k)}      
      \ar`[rd]`[l]`[dlll]`[d][dll] & 
      \\
      & 
      R^1 \beta_{2\ast} \big( I_9 \big)
      \ar[r] &
      \OsheafP(-1)
      \ar[r] &
      0
      \ar[r] &
      0
      \,.
    }}
\end{equation}
The restriction map $r$ works as follows. It takes a local function
$f$ in a neighborhood of $\beta_2(p_k) \in \CP^1$, and pulls it back
to a local function on $B_2$. This function is then restricted to the
three points $p_k$, $p_{3+k}$, and $p_{6+k}$. Since all functions on
an elliptic fiber are constant, the restriction to these three points
yields the same value. Hence, the image of $r$ is one dimensional
inside the three dimensional vector space over $\beta_2(p_k)$.  More
precisely, let
\begin{equation}
  G_1 \eqdef \{ e, g_1, g_1^2 \}
  \,,\quad
  G_2 \eqdef \{ e, g_2, g_2^2 \}  
\end{equation}
be the $\Z_3$ groups generated by $g_1$ and $g_2$. Then the $3$
dimensional stalk is the regular representation of $G_2$,
\begin{equation}
  \Reg(G_2) = 1 \oplus \chi_2 \oplus \chi_2^2
  \,,
\end{equation}
and the image of $r$ is the trivial representation. Knowing the
restriction map $r$ determines the pushdown of $I_9$ to be
\begin{equation}
  \begin{split}
    \beta_{2\ast}\big( I_9 \big) =&\, 
    \OsheafP(-3)
    \,,
    \\
    R^1\beta_{2\ast} \big( I_9 \big) =&\, 
    \OsheafP(-1)
    \oplus \left[
    \bigoplus_{k=1}^3 
    \big( \chi_2 \oplus \chi_2^2 \big) \Osheaf_{\beta_2(p_k)}
    \right]
    \,.
  \end{split}
\end{equation}

Now we can calculate the pushdown of $\chi_2\Wsheaf$ to the base
$\CP^1$. The vector bundle is defined via an extension
\begin{equation}
  \label{eq:Wbundlechi2}
  0 
  \longrightarrow
  \chi_2 \OsheafBtwo(-2 f) 
  \longrightarrow
  \chi_2 \Wsheaf
  \longrightarrow
  \chi_2^2
  \OsheafBtwo(2 f) \otimes I_9
  \longrightarrow
  0
  \,,
\end{equation}
and the associated long exact sequence of pushdowns is
\begin{equation}
  \label{eq:Wles}
  \vcenter{\xymatrix@R=10pt@M=4pt@H+=22pt{
      0 \ar[r] & 
      \chi_2 \OsheafP(-2)
      \ar[r] &
      \beta_{2\ast} \big( \chi_2 \Wsheaf \big)
      \ar[r] &
      \chi_2^2 \OsheafP(2) \otimes \beta_{2\ast} \big(I_9\big)
      \ar`[rd]^<>(0.5){\delta}`[l]`[dlll]`[d][dll] & 
      \\
      & 
      \chi_2 \OsheafP(-3)
      \ar[r] &
      R^1\beta_{2\ast} \big( \chi_2 \Wsheaf \big)
      \ar[r] &
      \chi_2^2 \OsheafP(2) \otimes R^1\beta_{2\ast} \big(I_9\big)
      \ar[r] &
      0
      \,.
    }}
\end{equation}
Since there are only maps of line bundles $\OsheafP(n)\to\OsheafP(m)$
for $n\leq m$, the coboundary map
\begin{equation}
  \chi_2^2 \OsheafP(2) \otimes \beta_{2\ast} \big(I_9\big)
  =
  \chi_2^2 \OsheafP(-1)
  \quad
  \stackrel{\delta}{\longrightarrow}
  \quad
  \chi_2 \OsheafP(-3)  
\end{equation}
has to be zero and the long exact sequence splits. There is no
extension ambiguity for the direct image, and we obtain
\begin{equation}
  \label{eq:beta2Wsheaf}
  \beta_{2\ast} \big( \chi_2 \Wsheaf \big)
  = 
  \chi_2 \OsheafP(-2)
  \oplus
  \chi_2^2 \OsheafP(-1)
  \,,
\end{equation}
whereas the derived direct image 
\begin{equation}
\begin{gathered}
  \label{eq:R1beta2Wsheafses}
  0
  \longrightarrow
  \chi_2 \OsheafP(-3)
  \longrightarrow
  R^1\beta_{2\ast} \big( \chi_2 \Wsheaf \big)
  \longrightarrow
  \hspace{4cm} \\ \hspace{4cm}
  \longrightarrow
  \chi_2^2 \OsheafP(1)
  \oplus \left[
    \bigoplus_{k=1}^3 
    \big( 1 \oplus \chi_2 \big) \Osheaf_{\beta_2(p_k)}
  \right]
  \longrightarrow
  0
\end{gathered}
\end{equation}
is not uniquely determined. To disentangle the short exact sequence,
we compare with relative duality. For that, we need the pushdown of
$(\chi_2\Wsheaf)^\dual$, which we compute from the short exact
eq.~\eqref{eq:Wdualbundle}. Alternatively, we can observe that
$\chi_2\Wsheaf$ is self-dual while $\Wsheaf$ alone is not, see
Footnote~\ref{note:Wdual}. Either way, one finds
\begin{equation}
  \beta_{2\ast} \Big[ \big( \chi_2 \Wsheaf \big)^\dual \Big] =
  \beta_{2\ast} \big( \chi_2 \Wsheaf \big)
  \,.  
\end{equation}
Using relative duality, this implies
\begin{equation}
\begin{gathered}
  \beta_{2\ast} \Big[ \big( \chi_2 \Wsheaf \big)^\dual \Big] =
  \Big( R^1\beta_{2\ast} \big( 
    \chi_2 \Wsheaf  \otimes K_{B_2|\CP^1}\big)  \Big)^\vee  = 
  \Big( R^1\beta_{2\ast} \big( 
    \chi_2 \Wsheaf \big)  \Big)^\vee 
  \otimes \OsheafP(-1)  
  \\  
  \Leftrightarrow  \quad
  \Big( R^1\beta_{2\ast} \big( 
  \chi_2 \Wsheaf \big)  \Big)^\vee 
  =
  \chi_2 \OsheafP(-1)
  \oplus
  \chi_2^2 \OsheafP
  \\  
  \Leftrightarrow  \quad
  \Big( R^1\beta_{2\ast} \big( 
  \chi_2 \Wsheaf \big)  \Big)^{\vee\vee}
  =
  \chi_2^2 \OsheafP(1)
  \oplus
  \chi_2 \OsheafP
  \,.
\end{gathered}
\end{equation}
Now the dual of the dual is not quite the original sheaf, since the
dual of a skyscraper sheaf is zero. So we can only conclude that 
\begin{equation}
  R^1\beta_{2\ast} \big( 
  \chi_2 \Wsheaf \big)  
  =
  \chi_2^2 \OsheafP(1)
  \oplus
  \chi_2 \OsheafP
  \oplus \text{torsion}
\end{equation}
However, together with the short exact sequence
eq.~\eqref{eq:R1beta2Wsheafses} this singles out the unique extension
\begin{equation}
  \label{eq:R1beta2Wsheaf}
  R^1\beta_{2\ast} \big( 
  \chi_2 \Wsheaf \big)  
  =
  \chi_2^2 \OsheafP(1)
  \oplus
  \chi_2 \OsheafP
  \oplus \left[
    \bigoplus_{k=1}^3 
    \Osheaf_{\beta_2(p_k)}
  \right]
  \,.
\end{equation}

\subsubsection{The Cohomology}

Now we have everything in place to compute the cohomology of
$\Vsheaf_1\otimes \Vsheaf_2$. We start with the \LSS{} pushing down to
$B_2$,
\begin{equation}
\label{eq:V1V2ss}
\begin{split}
  E_2^{p,q}\big(\Xt,\Vsheaf_1\otimes \Vsheaf_2\big) =&\, 
  H^p\Big( B_2,\, R^q\pi_{2\ast}\big( 
  \Vsheaf_1\otimes \Vsheaf_2
  \big) \Big)
  = 
  \\ =&\,
  H^p\Big( B_2,\, 
  2 \chi_2 \Wsheaf \otimes R^q\pi_{2\ast} \OsheafXt
  \Big)
  = 
  \\ =&\,
  \begin{cases}
    H^p\Big( B_2,\, 
    2 \chi_2 \Wsheaf 
    \otimes \pi_{2\ast} K_{\Xt|B_2}^\dual
    \Big)
    \,, & q=1
    \\[1ex]
    H^p\Big( B_2,\, 2 \chi_2 \Wsheaf \Big) 
    \,, & q=0 
    \,.
  \end{cases}
\end{split}
\end{equation}
We proceed by computing the cohomology of $\chi_2 \Wsheaf$ using
another \LSS,
\begin{equation}
  E_2^{p,q}\Big(B_2,\, \chi_2 \Wsheaf \Big) = 
  H^p\Big(\CP^1,\, R^q\beta_{2\ast}  
  \big( \chi_2 \Wsheaf \big)
  \Big) 
  \,.
\end{equation}
We computed the pushdowns in eqs.~\eqref{eq:beta2Wsheaf},
\eqref{eq:R1beta2Wsheaf}. The $q=0$ cohomology groups are
\begin{align}
  H^0\Big( \CP^1,\, \beta_{2\ast}  
  \big( \chi_2 \Wsheaf \big) \Big) \,&=
  H^0\Big( \CP^1,\,
  \chi_2 \OsheafP(-2)
  \oplus
  \chi_2^2 \OsheafP(-1)
  \Big)
  =0
  \,,
  \\
  H^1\Big( \CP^1,\, \beta_{2\ast}  
  \big( \chi_2 \Wsheaf \big) \Big) \,&=
  H^1\Big( \CP^1,\,
  \chi_2 \OsheafP(-2)
  \Big) 
  \oplus
  H^1\Big( \CP^1,\,
  \chi_2^2 \OsheafP(-1)
  \Big) 
  = \notag \\ &=
  H^0\Big( \CP^1,\,
  \big( \chi_2 \OsheafP(-2) \big)^\dual 
  \otimes K_{\CP^1}
  \Big)^\dual 
  \oplus
  0
  = \notag \\ &=
  H^0\Big( \CP^1,\,
  \big( \chi_2^2 \OsheafP(2) \big) 
  \otimes 
  \big( \chi_1 \otimes \OsheafP(-2) \big)
  \Big)^\dual 
  = \notag \\ &=
  \Big( \chi_1 \chi_2^2 \Big)^\dual
  =
  \chi_1^2 \chi_2
  \,.
\end{align}
For the $q=1$ terms, note that $G_1$ cyclically permutes the
skyscraper sheaves, so it acts in the regular representation on the
global sections of $\oplus_{k=1}^3 \Osheaf_{\beta_2(p_k)}$. One
obtains
\begin{multline}
  H^0\Big( \CP^1,\, R^1\beta_{2\ast}  
  \big( \chi_2 \Wsheaf \big) \Big)   
  = \\ =
  H^0\Big( \CP^1,\, 
    \chi_2^2 \OsheafP(1)
  \Big)
  \oplus
  H^0\Big( \CP^1,\, 
    \chi_2 \OsheafP
  \Big)
  \oplus 
  H^0\left( \CP^1,\, 
    \bigoplus_{k=1}^3 
    \Osheaf_{\beta_2(p_k)}
  \right)  
  = \\ =
  \Big( \chi_2^2 \big( 1 \oplus \chi_1 \big) \Big) \oplus
  \Big( \chi_2 \Big) \oplus
  \Big( \Reg(G_1) \Big) 
  = \\ =
  \chi_2^2 \oplus 
  \chi_1\chi_2^2 \oplus 
  \chi_2 \oplus
  1 \oplus 
  \chi_1 \oplus 
  \chi_1^2
\end{multline}
and
\begin{equation}
  H^1\Big( \CP^1,\, R^1\beta_{2\ast}  
  \big( \chi_2 \Wsheaf \big) \Big) =
  H^1\left( \CP^1,\, 
    \chi_2^2 \OsheafP(1)
    \oplus
    \chi_2 \OsheafP
    \oplus \left[
      \bigoplus_{k=1}^3 
      \Osheaf_{\beta_2(p_k)}
    \right]
  \right)
  = 0
  \,.
\end{equation}
Hence the \LSS{} for the $B_2\to\CP^1$ pushdown is
\begin{equation}
  E_2^{p,q}\Big(B_2,\, \chi_2 \Wsheaf \Big) = 
  \vcenter{\xymatrix@=1mm{
      {\scriptstyle q=1}\hspace{1.6mm} &
      \chi_2^2 \oplus 
      \chi_1\chi_2^2 \oplus 
      \chi_2 \oplus
      1 \oplus 
      \chi_1 \oplus 
      \chi_1^2
       & 0 \\
      {\scriptstyle q=0}\hspace{1.6mm} &
      0 & \chi_1^2\chi_2 \\
      \ar[]+/r 3.4mm/+/u 1.7mm/;[rr]+/r 3mm/+/r 3.4mm/+/u 1.7mm/
      \ar[]+/r 3.4mm/+/u 1.7mm/;[uu]+/u  2mm/+/r 3.4mm/+/u 1.7mm/
      & 
      {\vbox{\vspace{3.5mm}}\scriptstyle p=0} & 
      {\vbox{\vspace{3.5mm}}\scriptstyle p=1} 
    }}
\end{equation}
and we obtain
\begin{equation}
  H^p\Big(B_2,\, \chi_2 \Wsheaf \Big) =
  \begin{cases}
    0
    \,, & p=2 \\
      1 \oplus 
      \chi_1 \oplus 
      \chi_2 \oplus
      \chi_1^2 \oplus
      \chi_2^2 \oplus 
      \chi_1\chi_2^2 \oplus 
      \chi_1^2\chi_2 
    \,, & p=1 \\    
    0
    \,, & p=0\,. \\
  \end{cases}
\end{equation}
Similarly to $\chi_2\Wsheaf$, one can also compute the cohomology of 
\begin{equation}
  \chi_2\Wsheaf \otimes \pi_{2\ast} K_{\Xt|B_2}^\dual 
  =
  \chi_2\Wsheaf \otimes \Big( \chi_1^2 \OsheafXt(\phi) \Big)^\dual 
  =
  \chi_1 \chi_2 \Wsheaf \otimes \OsheafBtwo(-f)
\end{equation}
by yet another \LSS{}. However, there is a faster way to do so. We
already know that the \LSS{} on $\Xt$, eq.~\eqref{eq:V1V2ss},
degenerates because $E_2^{2,0}\big(\Xt,\Vsheaf_1\otimes
\Vsheaf_2\big)=0$. But this spectral sequence also has to yield the
fact\footnote{Group theory tells us that
  $\wedge^2\Rep{4}=\Rep{6}=\wedge^2\barRep{4} \in R[SU(4)]$.
  Therefore, $\wedge^2\Vsheaf=\wedge^2\Vsheaf^\dual$ and Serre duality
  forces the vanishing of the index.} that the (character-valued)
index $\idx\big(\Vsheaf_1\otimes \Vsheaf_2\big)=0$, and the only way
to accomplish that now is if the $q=0$ and $q=1$ terms coincide.
Hence, we conclude that
\begin{equation}
  H^p\Big( B_2,\, 
  2 \chi_2 \Wsheaf 
  \otimes \pi_{2\ast} K_{\Xt|B_2}^\dual
  \Big)
  =
  H^p\Big( B_2,\, 2 \chi_2 \Wsheaf \Big) 
  \,.
\end{equation}
Putting everything together, the $E_2^{p,q}\big(\Xt,\Vsheaf_1\otimes
\Vsheaf_2\big)=E_\infty^{p,q}\big(\Xt,\Vsheaf_1\otimes \Vsheaf_2\big)$
tableau is 
\begin{equation}
  \label{eq:V1V2ssend}
  \vcenter{\xymatrix@=3mm{
      {\scriptstyle q=1}\hspace{1.6mm} &
      0 &
      2 \oplus 
      2\chi_1 \oplus 
      2\chi_2 \oplus
      2\chi_1^2 \oplus
      2\chi_2^2 \oplus 
      2\chi_1\chi_2^2 \oplus 
      2\chi_1^2\chi_2 
      & 0 \\
      {\scriptstyle q=0}\hspace{1.6mm} &
      0 &
      2 \oplus 
      2\chi_1 \oplus 
      2\chi_2 \oplus
      2\chi_1^2 \oplus
      2\chi_2^2 \oplus 
      2\chi_1\chi_2^2 \oplus 
      2\chi_1^2\chi_2 
      & 0 \\
      \ar[]+/r 3.4mm/+/u 1.7mm/;[rrr]+/r 3mm/+/r 3.4mm/+/u 1.7mm/
      \ar[]+/r 3.4mm/+/u 1.7mm/;[uu]+/u  2mm/+/r 3.4mm/+/u 1.7mm/
      & 
      {\vbox{\vspace{3.5mm}}\scriptstyle p=0} & 
      {\vbox{\vspace{3.5mm}}\scriptstyle p=1} &  
      {\vbox{\vspace{3.5mm}}\scriptstyle p=2} 
    }}
\end{equation}
and, therefore, the desired cohomology group is
\begin{equation}
\label{eq:wedge2Vcoh}
\begin{split}
  H^p\Big(\Xt,\, \wedge^2\Vsheaf \Big) \,&=
  H^p\Big(\Xt,\, \Vsheaf_1 \otimes \Vsheaf_2 \Big) =
  \\ &=
  \begin{cases}
    0
    \,, & p=3 \\
      2 \oplus 
      2\chi_1 \oplus 
      2\chi_2 \oplus
      2\chi_1^2 \oplus
      2\chi_2^2 \oplus 
      2\chi_1\chi_2^2 \oplus 
      2\chi_1^2\chi_2 
    \,, & p=2 \\
      2 \oplus 
      2\chi_1 \oplus 
      2\chi_2 \oplus
      2\chi_1^2 \oplus
      2\chi_2^2 \oplus 
      2\chi_1\chi_2^2 \oplus 
      2\chi_1^2\chi_2 
    \,, & p=1 \\    
    0
    \,, & p=0\,. \\
  \end{cases}    
\end{split}
\end{equation}

\subsection{Existence of Extensions}
\label{sec:Vext}

In the definition of $\Vsheaf$, we assumed the existence of a generic
extension in two places. The first was the definition of the vector
bundle $\Wsheaf$ on $B_2$ in eq.~\eqref{eq:Wbundle}. There, the
Cayley-Bacharach theorem assured us that $\Wsheaf$ really is a vector
bundle, and not just a sheaf, if only we pick a generic nonzero
extension class. But, of course, we can only do so if there are any
nontrivial extensions. We can easily compute the space of extensions
using Serre duality and the \LSS,
\begin{equation}
\label{eq:ExtWresult}
\begin{split}
  \Ext^1&\Big( 
  \chi_2 \OsheafBtwo(2f) \otimes I_9
  ,\,
  \OsheafBtwo(-2f) \Big)
  =
  \Ext^1\Big( 
  \chi_1 \chi_2 \OsheafBtwo(3f) \otimes I_9
  ,\,
  K_{B_2} \Big)
  = \\ &=
  H^1\Big( B_2,\, 
  \chi_1 \chi_2 \OsheafBtwo(3f) \otimes I_9
  \Big)^\dual
  =
  H^0\Big( \CP^1,\, 
  \chi_1 \chi_2 \OsheafP(3) \otimes R^1\beta_{2\ast}\big(I_9\big)
  \Big)^\dual
  = \\ &=
  H^0\Big( \CP^1,\, 
  \chi_1 \chi_2 \OsheafP(2)
  \Big)^\dual  
  \oplus
  H^0\Big( \CP^1,\, 
  \bigoplus_{k=1}^3 \big( \chi_1 \chi_2^2 \oplus \chi_1 \big)
  \Osheaf_{\beta_2(p_k)}
  \Big)^\dual  
  = \\ &=
  \Big( 
  \chi_1\chi_2 \oplus 
  \chi_1^2\chi_2 \oplus 
  \chi_2
  \Big)^\dual 
  \oplus
  \Big( 
  (\chi_1\chi_2^2 \oplus \chi_1)\otimes  \Reg(G_1) 
  \Big)^\dual   
  = \\ &=
  \Reg(G_1 \times G_2) = \Reg(G)
  \,.
\end{split}
\end{equation}
The regular representation contains every $G\simeq \ZZZ$ character, so
in particular, there is a one-dimensional invariant subspace
\begin{equation}
  \Ext^1\Big( 
  \chi_2 \OsheafBtwo(2f) \otimes I_9
  ,\,
  \OsheafBtwo(-2f) \Big)^G  
  = 1
  \,.
\end{equation}
Using this extension, we can conclude that $\Wsheaf$ is a vector
bundle.

The second place where we assumed the existence of a nontrivial
extension was in the definition of $\Vsheaf$ itself in
eq.~\eqref{eq:Vbundle}. The vector bundle $\Vsheaf$ can only be stable
if the extension is nontrivial. Hence, we must ensure that there are,
indeed, nonzero elements of
\begin{equation}
\label{eq:Ext1V2V1}
\begin{split}    
  \Ext^1\Big( \Vsheaf_2,\, \Vsheaf_1 \Big) 
  \,&=
  H^1\Big(\Xt,\, \Vsheaf_1 \otimes \Vsheaf_2^\dual \Big)
  = \\ &=
  H^1\Big(\Xt,\, 
  2 \chi_2 \OsheafXt(-2\tau_1+2\tau_2) \otimes \pi_2^\ast(\Wsheaf)
  \Big)
  = \\ &=
  H^0\Big(B_2,\, 
  2 \chi_2 R^1\pi_{2\ast}\big(\OsheafXt(-2\tau_1)\big) 
  \otimes \OsheafBtwo(2t) \otimes \Wsheaf
  \Big)  
  = \\ &=
  H^0\Big(B_2,\, 
  12 \chi_2 
  \otimes
  \OsheafBtwo(2t-f) \otimes \Wsheaf
  \Big)  
  \,.
\end{split}
\end{equation}
We want to compute this cohomology group by pushing down to the base
$\CP^1$, for which we need the direct image of $\OsheafBtwo(2t-f)
\otimes \Wsheaf$. By twisting the short exact sequence
eq.~\eqref{eq:Wbundle}, we obtain
\begin{equation}
  0 
  \longrightarrow
  \OsheafBtwo(2t-3f)
  \longrightarrow
  \OsheafBtwo(2t-f) \otimes \Wsheaf
  \longrightarrow
  \chi_2
  \OsheafBtwo(2t+f)
  \otimes I_9
  \longrightarrow
  0
  \,.
\end{equation}
Because $R^1\beta_{2\ast}\OsheafBtwo(2t-3f)=0$ (for degree reasons)
the long exact sequence of pushdowns truncates and we obtain another
short exact sequence
\begin{equation}
\label{eq:O2tftimesWses}
\begin{split}
  0 
  \longrightarrow
  6 \OsheafP(-3)
  \longrightarrow
  \beta_{2\ast}\Big( \OsheafBtwo(2t-f) \otimes \Wsheaf \Big)
  \longrightarrow \hspace{3cm}
  \\
  \longrightarrow
  \chi_2
  \OsheafP(f)
  \otimes 
  \beta_{2\ast}\Big(\OsheafBtwo(2t)\otimes I_9\Big)
  \longrightarrow
  0
  \,.  
\end{split}
\end{equation}
To make use of this short exact sequence we first have to compute the
pushdown in the rightmost term. Using the definition of the ideal
sheaf, eq.~\eqref{eq:I9ses}, we obtain
\begin{equation}
\label{eq:O2tI9ses}
  \begin{split}
  0
  \longrightarrow
  \OsheafBtwo(2t)\otimes I_9
  \longrightarrow&\,
  \OsheafBtwo(2t)
  \longrightarrow
  \bigoplus_{i=1}^{9} \Osheaf_{p_i}
  \longrightarrow
  0
  \\ 
  &\Downarrow \, \scriptstyle{ \beta_{2\ast} }
  \\ 
  0
  \longrightarrow
  \beta_{2\ast}\Big(\OsheafBtwo(2t)\otimes I_9\Big)
  \longrightarrow&\,
  \underbrace{
    \beta_{2\ast}\Big(\OsheafBtwo(2t)\Big)
  }_{= 6 \OsheafP}
  \longrightarrow
  \bigoplus_{k=1}^{3} 3 \Osheaf_{\beta_2(p_k)}
  \longrightarrow
  0
  \,.  
  \end{split}
\end{equation}
Keeping in mind that really the Heisenberg group has to act on each
term, this leaves us with two possibilities: 
\begin{equation}
  \beta_{2\ast}\Big(\OsheafBtwo(2t)\otimes I_9\Big)
  =
  \begin{cases}
    3 \OsheafP(-3) \oplus 3 \OsheafP \\
    \qquad\text{or}\\
    3 \OsheafP(-2) \oplus 3 \OsheafP(-1)
    \,. \\
  \end{cases}
\end{equation}
Figuring out which possibility is realized turns out to be difficult
and lengthy. Since extensions in eq.~\eqref{eq:Vbundle} exist in any
case, we just remark that 
\begin{equation}
  \beta_{2\ast}\Big(\OsheafBtwo(2t)\otimes I_9\Big)
  =
  3 \OsheafP(-2) \oplus 3 \OsheafP(-1)
  \,.
\end{equation}
Second, we have to resolve any extension ambiguities in the above
short exact sequence eq.~\eqref{eq:O2tftimesWses}. We show that it
splits in Appendix~\ref{sec:pushdownext} and, hence, obtain that
\begin{equation}
  \beta_{2\ast}\Big( \OsheafBtwo(2t-f) \otimes \Wsheaf \Big)
  = 
  6 \OsheafP(-3)
  \oplus
  3 \OsheafP(-1) 
  \oplus 
  3 \OsheafP  
\end{equation}
and
\begin{equation}
  \dim_\C
  H^0\Big(B_2,\, 
  12
  \OsheafBtwo(2t-f) \otimes \Wsheaf
  \Big)
  = 
  36\,.
\end{equation}
Now we were not quite careful with the $G=\ZZZ$ group action. One has
to keep in mind that sections of $\OsheafBtwo(t)$ form representations
of the Heisenberg group. One can then easily show that the $36$
dimensional vector space eq.~\eqref{eq:Ext1V2V1} decomposes as
\begin{equation}
  \Ext^1\Big( \Vsheaf_2,\, \Vsheaf_1 \Big) 
  =
  4 \Reg\big(\ZZZ\big)
  \quad \Rightarrow \quad
  \Ext^1\Big( \Vsheaf_2,\, \Vsheaf_1 \Big)^G 
  = 4
  \,.
\end{equation}
We conclude that there are indeed equivariant extensions in all the
places where we assumed their existence.

\subsection{Vanishing of the First Chern Class}
\label{sec:c1vanish}

In our construction, we picked an $SU(4)$ subgroup of the $E_8$ gauge
bundle. This is not strictly necessary and one can also work with
$U(1)$ subgroups, as in~\cite{Blumenhagen:2005ga}. However, we will
only consider simple subgroups in this paper. Using the chosen
embedding, a $SU(4)$ principal bundle then gives rise to the desired
$E_8$ bundle. Of course, we are really working with a holomorphic
rank $4$ vector bundle and its $SL(4,\C)$ structure group, and then
use the deformation retract $SU(4)\to SL(4,\C)$.

But here it is important that the holomorphic vector bundle does have
a $SL(4,\C)$ structure group instead of the most general $GL(4,\C)$
structure group. Topologically, this manifests itself in the vanishing
of the first Chern class. Now, in constructing our vector bundle
$\Vsheaf$ we worked on the universal covering space $\Xt$, whereas we
should have worked on the \CY{} threefold $X=\Xt/G$. Of course, the
trace of the curvature of $\Vsheaf$ is the same as on $\Vsheaf/G$, so
the de Rham representative $c_1(\Vsheaf/G)\in H^2(X,\R)$ stays zero.
But the first Chern class really lives in $H^2(X,\Z)$ and quotienting
by $G\simeq\ZZZ$ can generate a torsion part. So, in general, only
\begin{equation}
  c_1(\Vsheaf) = 0 ~\in H^2\big(\Xt,\Z\big)
  \quad \Rightarrow \quad
  c_1(\Vsheaf/G) ~\in H^2_\mathrm{tors}\big(X,\Z\big)  
  = \Z_3 \oplus \Z_3
\end{equation}
holds and we must check that $c_1(\Vsheaf/G)=0$ separately.

The easiest way to ensure the vanishing of the first Chern class of
$\Vsheaf/G$ is to find a trivialization of the determinant line bundle
$\wedge^4\Vsheaf/G$. This we can discuss on the covering space $\Xt$
where, by construction, $\wedge^4\Vsheaf=\chi\OsheafXt$ for some
character $\chi$. The quotient is then trivial if and only if this
character is the identity representation,
\begin{equation}
  \Big( \chi \OsheafXt \Big)/G = \Osheaf_X
  \quad \Leftrightarrow \quad
  \chi = 1
  \,.
\end{equation}
The determinant line bundle for our rank $4$ bundle $\Vsheaf$
eq.~\eqref{eq:Vbundle} is
\begin{equation}
\begin{split}
  \wedge^4 \Vsheaf \,&=
  \Big( \wedge^2 \Vsheaf_1 \Big) 
  \otimes 
  \Big( \wedge^2 \Vsheaf_2 \Big)
  = \\ &=
  \Big( \chi_2 \OsheafXt(-\tau_1+\tau_2) \Big)^{\otimes 2}
  \otimes
  \Bigg[
  \Big( 
  \OsheafXt(-\tau_1+\tau_2) \otimes 
  \pi_2^{\ast}\big( \OsheafBtwo(-2f) \big)
  \Big)
   \otimes 
   \\ & \qquad \otimes
  \Big( 
  \OsheafXt(-\tau_1+\tau_2) \otimes 
  \pi_2^{\ast}\big( \chi_2\OsheafBtwo(2f) \big)
  \Big)
  \Bigg]
  = \chi_2^3 \OsheafXt = \OsheafXt
\end{split}
\end{equation}
and, hence,
\begin{equation}
  c_1\Big( \Vsheaf / G \Big) = 0 ~\in H^2(X,\Z)
\end{equation}
as it should.

We remark that there is at least one other equivariant action on the
rank $4$ vector bundle $\Vsheaf$ which also leads to vanishing first
Chern class. This nicely illustrates the importance of the equivariant
actions, and we discuss it in more detail in
Appendix~\ref{sec:otherequiv}.

\section{The Low Energy Spectrum}
\label{sec:spectrum}

\subsection{\texorpdfstring{$\Spin(10)$}{Spin(10)} Gauge Theory}
\label{sec:gauge}

\begin{figure}[htbp]
  \centering
  \input{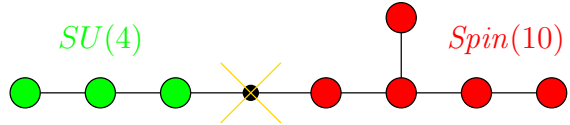}  
  \caption{Regular $SU(4)\times\Spin(10)$ subgroup of $E_8$.}
  \label{fig:E8spin10dynkin}
\end{figure}

First, let us only consider the effect of the $SU(4)$ instanton in the
visible $E_8$ gauge group. Then the $E_8$ gauge bosons acquire masses
except for the components which commute with the $SU(4)$. In other
words, the gauge group is broken to the commutant (or centralizer) of
$SU(4)\in E_8$. We pick a regular $SU(4)$ subgroup of $E_8$. Then, the
commutant can simply be read off from the extended Dynkin diagram, see
Figure~\ref{fig:E8spin10dynkin}.
The appearance of a $\Spin(10)$ gauge group is very desirable, since
one full generation of Standard Model matter quarks and leptons
(including a right-handed neutrino) fill out one $\Rep{16}$
representation of $\Spin(10)$.

The branching rule for the adjoint representation of $E_8$ is 
\begin{multline}
  \label{eq:E8Vbranch}
  R[E_8] \owns~
  \Rep{248} 
  = \\ =
  \big( \Rep{1},  \Rep{45} \big) \oplus
  \big( \Rep{15}, \Rep{1} \big) \oplus
  \big( \Rep{6}, \Rep{10} \big) \oplus
  \big( \Rep{4}, \Rep{16} \big) \oplus
  \big( \barRep{4}, \barRep{16} \big)
  ~\in R\big[ SU(4) \times \Spin(10) \big]
  \,.
\end{multline}
Correspondingly, the fermions in the adjoint of $E_8$ split into
fields charged only under $SU(4)$, only under $\Spin(10)$, or under
both groups. We identify the corresponding zero modes as
\begin{descriptionlist}
\item[$\big(\Rep{4},\Rep{16}\big)$:] The matter fields transforming
  in the $\Rep{16}$ of $\Spin(10)$. The number of such chiral
  multiplets is
  \begin{equation}
    H^1\Big( X,\, \Vsheaf/G \Big) = 
    H^1\Big( \Xt,\, \Vsheaf \Big)^G
    \,.
  \end{equation}
\item[$\big(\barRep{4},\barRep{16}\big)$:] Likewise, the number of
  $\barRep{16}$ matter fields. Similarly, their number is 
  \begin{equation}
    H^1\Big( X,\, \Vsheaf^\dual/G \Big) = 
    H^1\Big( \Xt,\, \Vsheaf^\dual \Big)^G
    \,.
  \end{equation}
\item[$\big(\Rep{6},\Rep{10}\big)$:] The matter fields transforming
  in the $\Rep{10}$ of $\Spin(10)$. Notice that it is a real
  representation, in particular
  \begin{equation}
    \wedge^2 \Rep{4}
    =
    \Rep{6}
    =
    \wedge^2 \barRep{4}
    ~\in R\big[SU(4)\big]
    \,.
  \end{equation}
  The number of chiral multiplets equals
  \begin{equation}
    H^1\Big( \Xt,\, \wedge^2 \Vsheaf \Big)^G    
    =
    H^1\Big( \Xt,\, \wedge^2 \Vsheaf^\dual \Big)^G    
    \,.
  \end{equation}
\item[$\big(\Rep{1},\Rep{45}\big)$:]
  The gauginos of the $\Spin(10)$ gauge group.
\item[$\big(\Rep{15},\Rep{1}\big)$:] The superpartners of the moduli
  fields (they are neutral under the $\Spin(10)$ gauge group). As
  \begin{equation}
    \Rep{15} = \Ad_{SU(4)} = 
    \Rep{4} \otimes \barRep{4} - \Rep{1} 
    ~\in R\big[SU(4)\big]
    \,,
  \end{equation}
  their number equals 
  \begin{equation}
    \bigg[
    H^1\Big( \Xt,\, \Vsheaf \otimes \Vsheaf^\dual \Big)
    - 
    \overbrace{H^1\Big( \Xt,\, \OsheafXt \Big)}^{=0}
    \bigg]^G
    =
    H^1\Big( \Xt,\, \Vsheaf \otimes \Vsheaf^\dual \Big)^G 
    \,.
  \end{equation}
\end{descriptionlist}

\subsection{Wilson Lines}
\label{sec:Wilson}

Of course, $\Spin(10)$ is not the right gauge group for
phenomenological purposes. It must be broken down to the Standard
Model $SU(3)_C \times SU(2)_L \times U(1)_Y$. Actually, to incorporate
the long lifetime of the nucleons, we postulate an extra $U(1)_{B-L}$
which naturally suppresses nucleon decay. Hence, we want to break 
\begin{equation}
  \Spin(10)
  \longrightarrow
  SU(3)_C \times SU(2)_L \times U(1)_Y \times U(1)_{B-L}
  \,.
\end{equation}
The obvious mechanism to do this is to make use of the $\pi_1(X)=\ZZZ$
fundamental group and add suitable Wilson lines. We showed
in~\cite{dP9Z3Z3} that there is a $\ZZZ\subset \Spin(10)$ subgroup
whose commutant is precisely $SU(3) \times SU(2) \times U(1)^2$.
Moreover, we computed the decomposition of the $\Rep{10}$ and
$\Rep{16}$ representations of $\Spin(10)$ under this $\ZZZ\times SU(3)
\times SU(2) \times U(1)^2$ subgroup and found\footnote{Here we are
  picking generators for the $U(1)^2$ which coincide with the
  conventional hypercharge and $B-L$. Our normalization is the one
  that gets rid of any fractions.}
\begin{equation}
  \label{eq:RepZZZsplit}
  \begin{split}
    \Rep{16} =&\,
    \chi_1^2\chi_2 
    \big( \Rep{3},    \Rep{2},  1,  1\big) \oplus
    \chi_1^2 
    \big( \Rep{1},    \Rep{1},  6,  3\big) \oplus
    \chi_1^2\chi_2^2 
    \big( \barRep{3}, \Rep{1}, -4, -1\big) \oplus
    \\ &\, \oplus
    \chi_2^2 
    \big( \barRep{3}, \Rep{1},  2, -1\big) \oplus
    \big( \Rep{1}, \barRep{2}, -3, -3\big) \oplus
    \chi_1 
    \big( \Rep{1},    \Rep{1},  0,  3\big) 
    \\ 
    \Rep{10} =&\,
    \chi_1 
    \big( \Rep{1},    \Rep{2},  3,  0\big) \oplus
    \chi_1\chi_2 
    \big( \Rep{3},    \Rep{1}, -2, -2\big) \oplus
    \\ &\, \oplus
    \chi_1^2 
    \big( \Rep{1}, \barRep{2}, -3,  0\big) \oplus
    \chi_1^2\chi_2^2 
    \big( \barRep{3}, \Rep{1},  2,  2\big)
    \,.
  \end{split} 
\end{equation}
This means the following in terms of vector bundles. On $\Xt$ we are
concerned with the rank $248$ vector bundle $\Esheaf_8^V$ which is
associated to the visible $E_8$ principal bundle via the adjoint
representation $\Rep{248}\in R[E_8]$. Thanks to the decomposition
eq.~\eqref{eq:E8Vbranch}, this vector bundle can be written in terms
of the $SU(4)$ vector bundle $\Vsheaf$ as
\begin{equation}
  \label{eq:E8Vbundle}
\begin{split}
  \Esheaf_8^V =&\, 
  \Big( \OsheafXt \otimes  \theta(\Rep{45}) \Big) \oplus
  \Big( \Ad(\Vsheaf) \otimes \theta(\Rep{1}) \Big) \oplus
  \\ &\,\oplus
  \Big( \wedge^2\Vsheaf \otimes \theta(\Rep{10}) \Big) \oplus
  \Big( \Vsheaf \otimes \theta(\Rep{16}) \Big) \oplus
  \Big( \Vsheaf^\dual \otimes \theta(\barRep{16}) \Big)
  \,.
\end{split}
\end{equation}
Here, the vector bundles $\theta(R)$ associated with a $\Spin(10)$
representation $R$ are just trivial rank $\dim(R)$ bundles on $\Xt$.
However, they inherit certain $\ZZZ$ actions from the representations
detailed in eq.~\eqref{eq:RepZZZsplit} and, therefore, their $\ZZZ$
quotient can be nontrivial.

\subsection{Matter Fields}
\label{sec:matter}

The matter fields in the low energy effective action are massless
relative to the compactification scale, and are computed as the index
of the Dirac operator coupled to the $\Esheaf_8^V$, see
eq.~\eqref{eq:E8Vbundle}. This index can be computed by the cohomology
groups of $\Esheaf_8^V$ which is what we are now going to do.

For example, let us focus on the $\wedge^2\Vsheaf\otimes
\theta(\Rep{10})$ summand. The vector bundle $\theta(\Rep{10})$
decomposes as
\begin{equation}
  \begin{split}
    \theta(\Rep{10}) =&\, 
    \chi_1 
    \theta\big( \Rep{1},    \Rep{2},  3,  0\big) \oplus
    \chi_1\chi_2 
    \theta\big( \Rep{3},    \Rep{1}, -2, -2\big) \oplus
    \\ &\, \oplus
    \chi_1^2 
    \theta\big( \Rep{1}, \barRep{2}, -3,  0\big) \oplus
    \chi_1^2\chi_2^2 
    \theta\big( \barRep{3}, \Rep{1},  2,  2\big)  
    \\ \simeq &\, 
    2 \chi_1 
    \OsheafXt
    \oplus
    3 \chi_1\chi_2 
    \OsheafXt
    \oplus
    2 \chi_1^2 
    \OsheafXt
    \oplus
    3 \chi_1^2\chi_2^2 
    \OsheafXt
    \,.
  \end{split}
\end{equation}
Therefore, the corresponding summand in eq.~\eqref{eq:E8Vbundle}
decomposes into
\begin{equation}
  \begin{split}
    \wedge^2 \Vsheaf \otimes \theta(\Rep{10}) =&\,
    \Big(
    \wedge^2\Vsheaf\otimes
    \chi_1 
    \theta\big( \Rep{1},    \Rep{2},  3,  0\big) 
    \Big) \oplus \Big(
    \wedge^2\Vsheaf\otimes
    \chi_1\chi_2 
    \theta\big( \Rep{3},    \Rep{1}, -2, -2\big)
    \Big) \oplus
    \\ &\, \oplus
    \Big(
    \wedge^2\Vsheaf\otimes
    \chi_1^2 
    \theta\big( \Rep{1}, \barRep{2}, -3,  0\big) 
    \Big) \oplus \Big(
    \wedge^2\Vsheaf\otimes
    \chi_1^2\chi_2^2 
    \theta\big( \barRep{3}, \Rep{1},  2,  2\big)
    \Big)
    \,.
  \end{split}  
\end{equation}
The $G=\ZZZ$ invariant part of the first cohomology group are the zero
modes. In our case, we obtain
\begin{equation}
  \begin{split}
    H^1\Big( \Xt,\,
    \wedge^2 \Vsheaf \otimes \theta(\Rep{10}) 
    \Big)^G
    =&\,
    \left[ 
      \chi_1\otimes
      H^1\Big( \Xt,\,
      \wedge^2\Vsheaf
      \Big)
    \right]^G \otimes
    \big( \Rep{1},    \Rep{2},  3,  0\big) 
    \,\oplus
    \\ &\, \oplus
    \left[ 
      \chi_1\chi_2 \otimes 
      H^1\Big( \Xt,\,    
      \wedge^2\Vsheaf
      \Big)
    \right]^G \otimes
    \big( \Rep{3},    \Rep{1}, -2, -2\big)
    \,\oplus
    \\ &\, \oplus
    \left[ 
      \chi_1^2 \otimes 
      H^1\Big( \Xt,\,
      \wedge^2\Vsheaf
      \Big)
    \right]^G \otimes
    \big( \Rep{1}, \barRep{2}, -3,  0\big) 
    \,\oplus
    \\ &\, \oplus
    \left[ 
      \chi_1^2\chi_2^2 \otimes 
      H^1\Big( \Xt,\,    
      \wedge^2\Vsheaf
      \Big)
    \right]^G \otimes
    \big( \barRep{3}, \Rep{1},  2,  2\big)
    \,.
  \end{split}  
\end{equation}
Hence $\big[ \chi_1\otimes H^1( \Xt,\, \wedge^2\Vsheaf )\big]^G$ is
the number of fields transforming in the representation $\big(
\Rep{1}, \Rep{2}, 3, 0\big)$, and so on. The same can be done for the
other matter fields coming from the $\Rep{16}$ and $\barRep{16}$ of
$\Spin(10)$.

The resulting spectrum can then be read off from the cohomology groups
which we computed in eqs.~\eqref{eq:Vcoh} and~\eqref{eq:wedge2Vcoh}.
We give the spectrum\footnote{To be precise, we only list the left
  chiral half, that is the number of $N=1$ left chiral multiplets. It
  is always understood that the particles are accompanied by their CPT
  conjugates.} in Table~\ref{tab:spectrum}, which is precisely three
families of quarks and leptons, together with two pairs of Higgs
doublets.
\begin{table}[htbp]
  \renewcommand{\arraystretch}{1.3}
  \centering
  \begin{tabular}{lccc}
    Multiplicity &
    Representation & 
    Name 
    \\ \hline
    $\displaystyle 
    3 = \left[ \chi_1^2\chi_2 \otimes
      H^1\Big( \Xt,\, \Vsheaf \Big) \right]^G 
    $ & 
    $\big( \Rep{3},    \Rep{2},  1,   1\big)$
    &
    left-handed quark
    \\
    $\displaystyle 
    3 = \left[ \chi_1^2 \otimes
      H^1\Big( \Xt,\, \Vsheaf \Big) \right]^G 
    $ & 
    $\big( \Rep{1},    \Rep{1}, 6,  3\big)$
    &
    left-handed charged anti-lepton
    \\
    $\displaystyle 
    3 = \left[ \chi_1^2 \chi_2^2 \otimes
      H^1\Big( \Xt,\, \Vsheaf \Big) \right]^G 
    $ & 
    $\big( \barRep{3}, \Rep{1}, -4, -1\big)$
    &
    left-handed anti-up
    \\
    $\displaystyle 
    3 = \left[ \chi_2^2 \otimes
      H^1\Big( \Xt,\, \Vsheaf \Big) \right]^G 
    $ & 
    $\big( \barRep{3}, \Rep{1},  2,  -1\big)$
    &
    left-handed anti-down
    \\
    $\displaystyle 
    3 = \left[ 
      H^1\Big( \Xt,\, \Vsheaf \Big) \right]^G 
    $ & 
    $\big( \Rep{1}, \barRep{2}, -3, -3\big)$
    &
    left-handed lepton
    \\
    $\displaystyle 
    3 = \left[ \chi_1 \otimes
      H^1\Big( \Xt,\, \Vsheaf \Big) \right]^G 
    $ & 
    $\big( \Rep{1},    \Rep{1}, 0,   3\big)$
    &
    left-handed anti-neutrino
    \\
    $\displaystyle 
    0 = \left[ \chi_1\chi_2^2 \otimes
      H^1\Big( \Xt,\, \Vsheaf^\dual \Big) \right]^G 
    $ & 
    $\big( \barRep{3},    \barRep{2}, -1,  -1\big)$
    &
    exotic
    \\
    $\displaystyle 
    0 = \left[ \chi_1 \otimes
      H^1\Big( \Xt,\, \Vsheaf^\dual \Big) \right]^G 
    $ & 
    $\big( \Rep{1},    \Rep{1}, -6,  -3\big)$
    &
    exotic
    \\
    $\displaystyle 
    0 = \left[ \chi_1 \chi_2 \otimes
      H^1\Big( \Xt,\, \Vsheaf^\dual \Big) \right]^G 
    $ & 
    $\big( \Rep{3}, \Rep{1},  4,  1\big)$
    &
    exotic
    \\
    $\displaystyle 
    0 = \left[ \chi_2 \otimes
      H^1\Big( \Xt,\, \Vsheaf^\dual \Big) \right]^G 
    $ & 
    $\big( \Rep{3}, \Rep{1},  -2,  1\big)$
    &
    exotic
    \\
    $\displaystyle 
    0 = \left[ 
      H^1\Big( \Xt,\, \Vsheaf^\dual \Big) \right]^G 
    $ & 
    $\big( \Rep{1}, \Rep{2}, 3, 3\big)$
    &
    exotic
    \\
    $\displaystyle 
    0 = \left[ \chi_1^2 \otimes
      H^1\Big( \Xt,\, \Vsheaf^\dual \Big) \right]^G 
    $ & 
    $\big( \Rep{1},    \Rep{1}, 0, -3\big)$
    &
    exotic
    \\ 
    $\displaystyle 
    2 = \left[ \chi_1 \otimes
      H^1\Big( \Xt,\, \wedge^2\Vsheaf \Big) \right]^G 
    $ & 
    $\big( \Rep{1},    \Rep{2}, 3, 0\big)$
    &
    up Higgs
    \\ 
    $\displaystyle 
    0 = \left[ \chi_1\chi_2 \otimes
      H^1\Big( \Xt,\, \wedge^2\Vsheaf \Big) \right]^G 
    $ & 
    $\big( \Rep{3},    \Rep{1}, -2, -2\big)$
    &
    exotic
    \\ 
    $\displaystyle 
    2 = \left[ \chi_1^2 \otimes
      H^1\Big( \Xt,\, \wedge^2\Vsheaf \Big) \right]^G 
    $ & 
    $\big( \Rep{1}, \barRep{2}, -3, 0\big)$
    &
    down Higgs
    \\ 
    $\displaystyle 
    0 = \left[ \chi_1^2\chi_2^2 \otimes
      H^1\Big( \Xt,\, \wedge^2\Vsheaf \Big) \right]^G 
    $ & 
    $\big( \barRep{3}, \Rep{1}, 2, 2\big)$
    &
    exotic
  \end{tabular}
  \caption{Low energy spectrum. Note that all exotic representations
    of the gauge group come with multiplicity zero.
  }
  \label{tab:spectrum}
\end{table}


\section{An Alternative Hidden Sector}
\label{sec:anomaly}

\subsection{Anomaly cancellation without Five-Branes}
\label{sec:withoutfivebranes}

Instead of using five-branes, we can also cancel the anomaly by adding
another $SU(2)$ instanton to the hidden $E_8$. In total, we then have
a $SU(2)\times SU(2)$ gauge bundle embedded into the hidden $E_8$. We
denote the additional hidden $SU(2)$ bundle by $\Ssheaf$ and define it
as a pullback from $B_1$, that is,
\begin{equation}
  \label{eq:Ssheafdef}
  \Ssheaf
  \eqdef 
  \pi_1^\ast\big( \Ssheaf_B \big)
  \,.
\end{equation}
The rank $2$ bundle $\Ssheaf_B$ on $B_1$ is defined as the generic
extension in the short exact sequence
\begin{equation}
  \label{eq:SBdef}
  0
  \longrightarrow
  \Osheaf(-2f)
  \longrightarrow
  \Ssheaf_B
  \longrightarrow
  \Osheaf(2f)
  \otimes I_6^f
  \longrightarrow
  0
  \,.
\end{equation}
The definition of the ideal sheaf $I_6^f$ is delicate, and we postpone
it for the moment. Its detailed definition is important for the
cohomology groups of $\Ssheaf$, but the intricacies are not detected
by the Chern classes. 

The second Chern class of the ideal sheaf $I_6^f$ on $B_1$ is simply
$6$, the number of points. Therefore,
\begin{equation}
  \label{eq:Schern}
  ch(\Ssheaf) = 
  e^{-2f} + e^{2f} \Big(1-6\tau_1^2\Big)
  = 2 - 6 \tau_1^2
  \,.
\end{equation}
We can immediately read off that
\begin{equation}
  c_2(\Ssheaf) = 6 \tau_1^2
  \,,
\end{equation}
which is precisely the Poincar\'e dual of the curve on which we had to
wrap the five-brane in the previous Section. This is exactly what we
want, since for two $SU(2)$ bundles the second Chern classes just
adds,
\begin{equation}
  c_2\Big( \Hsheaf \oplus \Ssheaf \Big) =
  c_2\big( \Hsheaf \big) + c_2\big( \Ssheaf \big)
  \,.
\end{equation}
Hence, we can use $\Hsheaf\oplus\Ssheaf$ as a hidden gauge instanton
and get an anomaly free model without any five-branes.
\begin{equation}
  \label{eq:TXVHSanomaly}
  c_2\big( T\Xt \big) 
  - c_2\big( \Vsheaf \big) 
  - c_2\big( \Hsheaf \oplus \Ssheaf \big)
  = 0
  \,.
\end{equation}

\subsection{The Ideal Sheaf}

As mentioned above, the ideal sheaf $I_6^f$ requires additional
discussion. So far, all ideal sheaves used a generic $G$ orbit, which
has $|G|=9$ points.  But on the $dP_9$ surface $B_1$ there are two
shorter orbits, both of length $3$. One comprises the three $g_1$
fixed points and the other the three $g_2$ fixed points. For the ideal
sheaf $I_6^f$, we take the three $g_2$ fixed points $p_1$, $p_2$, and
$p_3$ with multiplicity $2$.

The three fixed points $p_i$ are in three different $I_1$
fibers\footnote{Since $g_2$ is translation by a section, its fixed
  points necessarily lie in singular fibers.} of the elliptic
fibration. But the ideal sheaf is not uniquely determined by knowing
that there are three points of multiplicity two. Recall that the ideal
sheaf of points with multiplicity one is just the sheaf of analytic
functions vanishing at these points. Now, the multiplicity two means
that the function and a first derivative has to vanish at the three
point $p_1$, $p_2$, and $p_3$. But at any point there are two
independent first derivatives, since the $dP_9$ surface is two
dimensional. We are going to demand that the first derivative in the
fiber direction vanishes, and express that by the superscript ``$f$''
in $I_6^f$. The extension eq.~\eqref{eq:SBdef} still satisfies the
Cayley-Bacharach condition, since a global section of
$\OsheafBone(4f)\otimes K_{B_1}$ vanishing at a point in a fiber
vanishes identically on that fiber. Therefore, a generic extension in
the short exact sequence eq.~\eqref{eq:SBdef} is a rank $2$ vector
bundle.

In order to compute the cohomology groups of $\Ssheaf$, we need to
know the pushdown of $I_6^f$ to the base $\CP^1$. The ideal sheaf
fits into a short exact sequence
\begin{equation}
  \label{eq:I6fses}
  0
  \longrightarrow
  I_6^f
  \longrightarrow
  \OsheafBone
  \longrightarrow
  \bigoplus_{i=1}^{3} \Osheaf_{2 p_i}
  \longrightarrow
  0
\end{equation}
leading to a long exact sequence for the pushdown
\begin{equation}
  \label{eq:I6fles}
  \vcenter{\xymatrix@R=10pt@M=4pt@H+=22pt{
      0 \ar[r] & 
      \beta_{1\ast}\big( I_6^f \big)
      \ar[r] &
      \OsheafP 
      \ar[r]^<>(0.5){r} &
      \bigoplus_{i=1}^{3} \Osheaf_{2 \beta_1(p_i)}      
      \ar`[rd]`[l]`[dlll]`[d][dll] & 
      \\
      & 
      R^1 \beta_{1\ast} \big( I_6^f \big)
      \ar[r] &
      \chi_1 \OsheafP(-1)
      \ar[r] &
      0
      \ar[r] &
      0
      \,.
    }}
\end{equation}
Now the restriction map $r$ works as follows. Let $f$ be a local
section of $\OsheafP$, that is, a holomorphic function over a small
open set. Then this function is pulled back to $B_1$, that is, it is
taken to be constant along the fiber of the elliptic fibration
$\beta_1:B_1\to \CP^1$. Now restrict to the skyscraper sheaf $\oplus
\Osheaf_{2p_i}$.  If the open set does not contain $\beta_1(p_i)$,
then the restriction is simply zero. And if the open set does contain
$\beta_1(p_i)$, then the pullback function $f\circ \beta_1$ is
restricted to its value at $p_i$ and its first derivative in the fiber
direction at $p_i$. By construction, this first derivative is always
zero for a function pulled back from the base $\CP^1$.

To summarize, the fiber of the skyscraper sheaf $\oplus_{i=1}^{3}
\Osheaf_{2 \beta_1(p_i)}$ over $p_i$ is $\C^2$, and the image of the
restriction map is $\C\oplus \{0\}$. The generator $g_1$ acts in the
regular representation on the skyscraper sheaf, permuting the three
points $\beta_1(p_1)$, $\beta_1(p_2)$, and $\beta_1(p_3)$, while the
second generator $g_2$ acts as\footnote{The first derivative in the
  fiber direction picks up this subtle phase. However, we can afford
  to gloss over it because the torsion part of
  $R^1\pi_{1\ast}\big(I_6^f\big)$ is absorbed in a nontrivial
  extension, see eq.~\eqref{eq:SBtwosesR1pi}.} $1\oplus \chi_2$.
Hence we can split the long exact sequence into
\begin{equation}
  \vcenter{\xymatrix@R=1mm{
      0 \ar[r] & 
      \beta_{1\ast}\big( I_6^f \big)
      \ar[r] &
      \OsheafP 
      \ar[r] &
      \bigoplus_{i=1}^{3} \Osheaf_{\beta_1(p_i)}      
      \ar[r]
      & 0
      \,,
      \\
      0 \ar[r]
      & 
      \bigoplus_{i=1}^{3} \chi_2 \Osheaf_{\beta_1(p_i)}      
      \ar[r]
      &
      R^1 \beta_{1\ast} \big( I_6^f \big)
      \ar[r] &
      \chi_1 \OsheafP(-1)
      \ar[r] &
      0
      \,.
    }}
\end{equation}
These short exact sequences uniquely determine the pushdown of the
ideal sheaf to be
\begin{align}
  \beta_{1\ast}\big( I_6^f \big)
  &=
  \OsheafP(-3)
  \,,
  &
  R^1 \beta_{1\ast} \big( I_6^f \big)
  &=
  \bigoplus_{i=1}^{3} \chi_2 \Osheaf_{\beta_1(p_i)}      
  \oplus \chi_1 \OsheafP(-1)
  \,.
\end{align}

\subsection{Cohomology of \texorpdfstring{$\Ssheaf$}{S}}

To be able to apply the \LSS{}, we need to know the pushdown of
$\Ssheaf_B$ from $B_1$ to $\CP^1$. By definition, $\Ssheaf_B$ fits
into the short exact sequence eq.~\eqref{eq:SBdef}. Hence, we obtain
the following long exact sequence for the pushdown
\begin{equation}
  \vcenter{\xymatrix@R=10pt@M=4pt@H+=22pt{
      0 \ar[r] & 
      \OsheafP(-2)
      \ar[r] &
      \beta_{1\ast}\Ssheaf_B
      \ar[r] &
      \beta_{1\ast}\big( I_6^f \big) \otimes \OsheafP(2)
      \ar`[rd]^<>(0.5){\delta}`[l]`[dlll]`[d][dll] & 
      \\
      & 
      \chi_1 \OsheafP(-3)
      \ar[r] &
      R^1\beta_{1\ast}\Ssheaf_B
      \ar[r] &
      \Big(
        R^1 \beta_{1\ast} \big( I_6^f \big)
      \Big) \otimes \OsheafP(2)
      \ar[r] &
      0
      \,.
    }}  
\end{equation}
The coboundary map 
\begin{equation}
  \delta: \OsheafP(-1) \to \OsheafP(-3)
\end{equation}
is zero for degree reasons and, therefore, the long exact sequence
splits into two short exact sequences
\begin{subequations}
\begin{gather}
  \label{eq:SBtwosespi}
  \xymatrix{
    0 \ar[r] & 
    \OsheafP(-2)
    \ar[r] &
    \beta_{1\ast}\big( \Ssheaf_B \big)
    \ar[r] &
    \OsheafP(-1)
    \ar[r]
    & 0
  }
  \,,
  \\
  \label{eq:SBtwosesR1pi}
  \xymatrix{
    0 \ar[r] 
    &
    \chi_1 \OsheafP(-3)
    \ar[r]
    &
    R^1\beta_{1\ast}\big( \Ssheaf_B \big)
    \ar[r] &
    \bigoplus_{i=1}^{3} \chi_2 \Osheaf_{\beta_1(p_i)}      
    \oplus \chi_1 \OsheafP(1)
    \ar[r] &
    0
  }
  \,.
\end{gather}
\end{subequations}
The first short exact sequence uniquely determines
\begin{equation}
  \beta_{1\ast}\big( \Ssheaf_B \big)
  =
  \OsheafP(-2) \oplus \OsheafP(-1)  
  \,,
\end{equation}
but the extension in the second short exact sequence is not unique. We
fix the ambiguity by combining the fact that $\Rep{2}=\barRep{2}$ in
$SU(2)$, hence the bundle $\Ssheaf_B$ is isomorphic to its dual
$\Ssheaf_B^\vee$, and relative duality to get
\begin{equation}
\label{eq:SBR1pimodtorsion}
\begin{split}
  \beta_{1\ast} \big(\Ssheaf_B\big) &=
  \Big( R^1\beta_{1\ast} \big( 
    \Ssheaf_B \otimes K_{B_1|\CP^1}\big)  \Big)^\vee  = 
  \Big( R^1\beta_{1\ast} \big( 
    \Ssheaf_B \big)  \Big)^\vee 
  \otimes \chi_1 \OsheafP(-1)  
  \\  
  &
  \begin{split}
    \Rightarrow ~
    R^1\beta_{1\ast} \big( \Ssheaf_B \big) &=
    \Big( \beta_{1\ast} \big(\Ssheaf_B\big) \Big)^\vee 
    \otimes \chi_1 \OsheafP(-1)  
    \oplus \text{torsion}
    = \\ &=
    \chi_1 \OsheafP(1) \oplus \chi_1 \OsheafP  
    \oplus \text{torsion}      
    \,.
  \end{split}
\end{split}
\end{equation}
The only extension
in the short exact sequence eq.~\eqref{eq:SBtwosesR1pi} that also
satisfies eq.~\eqref{eq:SBR1pimodtorsion} is
\begin{equation}
  R^1 \beta_{1\ast}\big( \Ssheaf_B \big)
  =
  \chi_1 \OsheafP(1) \oplus \chi_1 \OsheafP  
  \,.
\end{equation}

Now it is a simple application of the \LSS{} to compute the cohomology
of $\Ssheaf$. In the first step, we push $\Ssheaf$ back down to $B_1$
and obtain
\begin{equation}
  E_2^{p,q}\big(\Xt,\Ssheaf\big) = 
  H^p\Big( B_1,\, R^q\pi_{1\ast}\big(\pi_1^\ast \Ssheaf_B\big) \Big)
  = 
  \begin{cases}
    H^p\Big( B_1,\, 
    R^1\pi_{1\ast}\big(\pi_1^\ast \Ssheaf_B\big) \Big)
    \,, & q=1
    \\[1ex]
    H^p\Big( B_1,\, \Ssheaf_B \Big) 
    \,, & q=0 
    \,.
  \end{cases}
\end{equation}
The cohomology groups of the two sheaves on $B_1$ can, in turn, again
be computed using another \LSS. Starting with $H^p( B_1,\,
\Ssheaf_B)$, one finds
\begin{equation}
  E_2^{p,q}\Big(B_1,\, \Ssheaf_B\Big) = 
  H^p\Big(\CP^1,\, R^q\beta_{1\ast}\Ssheaf_B\Big) = 
  \vcenter{\xymatrix@=1mm{
      {\scriptstyle q=1}\hspace{1.6mm} &
      2 \chi_1 + \chi_1^2 & 0 \\
      {\scriptstyle q=0}\hspace{1.6mm} &
      0 & \chi_1^2 \\
      \ar[]+/r 3.4mm/+/u 1.7mm/;[rr]+/r 3mm/+/r 3.4mm/+/u 1.7mm/
      \ar[]+/r 3.4mm/+/u 1.7mm/;[uu]+/u  2mm/+/r 3.4mm/+/u 1.7mm/
      & 
      {\vbox{\vspace{3.5mm}}\scriptstyle p=0} & 
      {\vbox{\vspace{3.5mm}}\scriptstyle p=1} 
    }}
  =
  E_\infty^{p,q}\Big(B_1, \Ssheaf_B\Big)
\end{equation}
Each entry in the $E_2$ tableau above can easily be computed using
eq.~\eqref{eq:P1cohomology}. For example, $E_2^{0,1}(B_1,\Ssheaf_B)$
is
\begin{equation}
\begin{split}
  H^1\Big( \CP^1,\, \beta_{1\ast} \Ssheaf_B \Big) &= 
  H^1\Big( \CP^1,\, \OsheafP(-2) \oplus \OsheafP(-1) \Big) 
  = \\ &=  
  H^0\Big( \CP^1,\, 
  \big(\OsheafP(-2) \oplus \OsheafP(-1)\big)^\vee
  \otimes K_{\CP^1}
  \Big)^\vee 
  = \\ &=  
  H^0\Big( \CP^1,\, 
  \chi_1 \OsheafP \oplus \chi_1 \OsheafP(-1)
  \Big)^\vee = \big( \chi_1 \big)^\vee = \chi_1^2
  \,.
\end{split}
\end{equation}
Next, we need the cohomology groups of $R^1\pi_{1\ast}\big(\pi_1^\ast
\Ssheaf_B\big)$. Using the projection formula and relative duality, we
identify
\begin{equation}
  R^1\pi_{1\ast}\big(\pi_1^\ast \Ssheaf_B\big) 
  =
  \Ssheaf_B \otimes
  R^1\pi_{1\ast}\big( \OsheafXt \big) 
  =
  \Ssheaf_B \otimes
  \big( \pi_{1\ast}K_{\Xt|B_1}\big)^\vee
  =
  \Ssheaf_B \otimes \OsheafBone(-f)
  \,.
\end{equation}
Therefore, the \LSS{} for pushing it down to the base $\CP^1$ is
\begin{multline}
  E_2^{p,q}\Big(B_1,\, 
  R^1\pi_{1\ast}\big(\pi_1^\ast \Ssheaf_B\big)  \Big)
  =
  H^p\Big(\CP^1,\, R^q\beta_{1\ast}\big(\Ssheaf_B\big) \otimes 
  \OsheafP(-1) \Big) 
  = \\ =
  \vcenter{\xymatrix@=1mm{
      {\scriptstyle q=1}\hspace{1.6mm} &
      \chi_1 & 0 \\
      {\scriptstyle q=0}\hspace{1.6mm} &
      0 & 2\chi_1^2+\chi_1 \\
      \ar[]+/r 3.4mm/+/u 1.7mm/;[rr]+/r 3mm/+/r 3.4mm/+/u 1.7mm/
      \ar[]+/r 3.4mm/+/u 1.7mm/;[uu]+/u  2mm/+/r 3.4mm/+/u 1.7mm/
      & 
      {\vbox{\vspace{3.5mm}}\scriptstyle p=0} & 
      {\vbox{\vspace{3.5mm}}\scriptstyle p=1} 
    }}
  =
  E_\infty^{p,q}\Big(B_1,\, 
  R^1\pi_{1\ast}\big(\pi_1^\ast \Ssheaf_B\big)  \Big)
  \,.
\end{multline}
We determined all entries of the $E_2^{p,q}(\Xt,\Ssheaf)$ tableau to
be 
\begin{equation}
  E_2^{p,q}\Big(\Xt,\,\Ssheaf\Big) = 
  \vcenter{\xymatrix@=3mm{
      {\scriptstyle q=1}\hspace{1.6mm} &
      0 \ar[]!/d 1mm/;[drr]!/u 1mm/^(0.85){d_3}
      & 2\chi_1+2\chi_1^2 & 0 \\
      {\scriptstyle q=0}\hspace{1.6mm} &
      0 & 2\chi_1+2\chi_1^2 & 0 \\
      \ar[]+/r 3.4mm/+/u 1.7mm/;[rrr]+/r 3mm/+/r 3.4mm/+/u 1.7mm/
      \ar[]+/r 3.4mm/+/u 1.7mm/;[uu]+/u  2mm/+/r 3.4mm/+/u 1.7mm/
      & 
      {\vbox{\vspace{3.5mm}}\scriptstyle p=0} & 
      {\vbox{\vspace{3.5mm}}\scriptstyle p=1} &  
      {\vbox{\vspace{3.5mm}}\scriptstyle p=2} 
    }}
  \,.
\end{equation}
The potential $d_3$ differential vanishes by dimension and the
spectral sequence collapses. Summing up the diagonals, we determine
the cohomology of $\Ssheaf$ to be 
\begin{equation}
  H^p\Big(\Xt,\,\Ssheaf\Big) =
  \begin{cases}
    0
    \,, & p=3 \\
    2\chi_1+2\chi_1^2
    \,, & p=2 \\
    2\chi_1+2\chi_1^2
    \,, & p=1 \\    
    0
    \,, & p=0\,. \\
  \end{cases}
\end{equation}
Note that there is no invariant part,
\begin{equation}
  \label{eq:Scohinv}
  H^\ast\Big(\Xt,\,\Ssheaf\Big)^G =  0
  \,.
\end{equation}

\subsection{Cohomology of \texorpdfstring{$\Hsheaf\otimes\Ssheaf$}{H
    tensor S}}

To compute the cohomology groups of $\Hsheaf\otimes\Ssheaf$, we again
utilize the \LSS. First, we push down to $B_1$. The short exact
sequence eq.~\eqref{eq:Hdef} tensored with $\Ssheaf$ yields
\begin{equation}
  \label{eq:HSdef}
  0 
  \longrightarrow
  \underbrace{
    \OsheafXt( 2 \tau_1 + \tau_2 - \phi)
    \otimes \Ssheaf
  }_{\mathclap{
      =
      \pi_1^\ast\left( 
      \OsheafBone(2t) \otimes \Ssheaf_B
      \right)
      \otimes 
      \pi_2^\ast
      \OsheafBtwo(t-f)
    }}
  \longrightarrow
  \Hsheaf
  \otimes \Ssheaf
  \longrightarrow
  \underbrace{
    \OsheafXt(-2\tau_1 - \tau_2 + \phi)  
    \otimes \Ssheaf
  }_{\mathclap{
      =
      \pi_1^\ast\left( 
      \OsheafBone(-2t) \otimes \Ssheaf_B
      \right)
      \otimes 
      \pi_2^\ast
      \OsheafBtwo(-t+f)
    }}
  \longrightarrow
  0
  \,.
\end{equation}
Now we push down to $B_1$, and from the associated long exact sequence
we can immediately read off
\begin{equation}
  \begin{aligned}
    R_1\pi_{1\ast} \big( \Hsheaf \otimes \Ssheaf \big)
    ~&=~
    \OsheafBone(-2t) \otimes \Ssheaf_B
    \otimes 
    \Big( 3  \OsheafBone \Big)    
    \\
    \pi_{1\ast} \big( \Hsheaf \otimes \Ssheaf \big)
    ~&=~
    \OsheafBone(2t) \otimes \Ssheaf_B
    \otimes 
    \Big( 3  \OsheafBone(-f) \Big)
    \,.
  \end{aligned}
\end{equation}
To compute the subsequent pushdown to the base $\CP^1$, we first have
to push down $I_6^f \otimes \OsheafBone(2t)$. This can be obtained
from the defining short exact sequence
\begin{equation}
\label{eq:O2tI6ses}
  \begin{split}
  0
  \longrightarrow
  \OsheafBone(2t)\otimes I_6^f
  \longrightarrow&\,
  \OsheafBone(2t)
  \longrightarrow
  \bigoplus_{i=1}^{3} \Osheaf_{2p_i}
  \longrightarrow
  0
  \\ 
  &\Downarrow \, \scriptstyle{ \beta_{1\ast} }
  \\ 
  0
  \longrightarrow
  \beta_{1\ast}\Big(\OsheafBone(2t)\otimes I_6^f\Big)
  \longrightarrow&\,
  \underbrace{
    \beta_{1\ast}\Big(\OsheafBtwo(2t)\Big)
  }_{= 6 \OsheafP}
  \longrightarrow
  \bigoplus_{i=1}^{3} \Osheaf_{2 \beta_1(p_i)}
  \longrightarrow
  0
  \,.  
  \end{split}
\end{equation}
Considering that a section of $\OsheafBone(2t)$ and its derivative do
not vanish simultaneously, we conclude that
\begin{equation}
  \beta_{1\ast}\Big( \OsheafBone(2t)\otimes I_6^f \Big) =
  6 \OsheafP(-1)
  \,,  \qquad 
  R^1\beta_{1\ast}\Big( \OsheafBone(2t)\otimes I_6^f \Big) =
  0
  \,.
\end{equation}
Tensoring the short exact sequence defining $S_B$,
eq.~\eqref{eq:SBdef}, with $\OsheafBone(2t)$ and pushing down one
obtains
\begin{equation}
  0
  \longrightarrow
  6 \OsheafP(-2)
  \longrightarrow
  \beta_{1\ast}\Big( \OsheafBone(2t)\otimes \Ssheaf_B \Big)
  \longrightarrow
  6 \OsheafP(1)
  \longrightarrow
  0
  \,.
\end{equation}
At this point, we assume that the extension is generic, that is, 
\begin{equation}
  \beta_{1\ast}\Big( \OsheafBone(2t)\otimes \Ssheaf_B \Big) =  
  6 \OsheafP(-1) \oplus 6 \OsheafP
  \,.
\end{equation}
By relative duality and the fact that $\Ssheaf_B = \Ssheaf_B^\dual$,
this implies
\begin{equation}
  \beta_{1\ast}\Big( \OsheafBone(-2t)\otimes \Ssheaf_B \Big) =  
  6 \OsheafP(-1) \oplus 6 \OsheafP
  \,.  
\end{equation}
Putting everything together, we can easily compute every entry in the
\LSS{}. The group action is necessarily the regular representation,
coming from the tensor product of Heisenberg group
representations. Therefore, one obtains
\begin{equation}
  \label{eq:SHcoh}
  H^p\Big(\Xt,\, \Hsheaf\otimes\Ssheaf \Big) =
  \begin{cases}
    0
    \,, & p=3 \\
    2 \Reg(\ZZZ)    
    \,, & p=2 \\
    2 \Reg(\ZZZ)
    \,, & p=1 \\    
    0
    \,, & p=0\,. \\
  \end{cases}
\end{equation}

\subsection{The Hidden Spectrum}

Having constructed the two $SU(2)$ bundles $\Hsheaf$ and $\Ssheaf$, we
must specify an embedding $SU(2)\times SU(2)\subset E_8$ to construct
the hidden $E_8$ gauge bundle. We will stick to the simplest
possibility and choose the maximal regular subgroup
\begin{equation}
  SU(2) \times SU(2) \times \Spin(12) \subset E_8
  \,.
\end{equation}
Hence, the hidden $E_8$ gauge groups is broken down to $\Spin(12)$.
The corresponding branching rule for the adjoint of $E_8$ is
\begin{equation}
  \Rep{248} = 
  \big( \Rep{1}, \Rep{1}, \Rep{66} \big) \oplus 
  \big( \Rep{3}, \Rep{1}, \Rep{1} \big) \oplus 
  \big( \Rep{1}, \Rep{3}, \Rep{1} \big) \oplus 
  \big( \Rep{1}, \Rep{2}, \Rep{32} \big) \oplus 
  \big( \Rep{2}, \Rep{1}, \Rep{32} \big) \oplus 
  \big( \Rep{2}, \Rep{2}, \Rep{12} \big)
  \,.
\end{equation}
The first three summands correspond to $\Spin(12)$ gauginos and moduli
for $\Hsheaf$ and $\Ssheaf$. Then there are potentially matter fields
corresponding to $H^1\big(X,\Hsheaf/G\big)$ and
$H^1\big(X,\Ssheaf/G\big)$. As we saw previously, there are no such
light matter fields by eqs.~\eqref{eq:Hcohvanish}
and~\eqref{eq:Scohinv}.

However, there is a final contribution to the spectrum consisting of
matter fields transforming in the $\Rep{12}$ of $\Spin(12)$. Their
multiplicity is
\begin{equation}
  H^1\Big( X,\, \big(\Hsheaf\otimes\Ssheaf\big)\big/G \Big) 
  = 
  H^1\Big( \Xt,\, \Hsheaf\otimes\Ssheaf \Big)^G
  = 2  
  \,,
\end{equation}
where we have used the cohomology groups computed in
eq.~\eqref{eq:SHcoh}. Note that because of the occurrence of the
regular representation of $G=\ZZZ$ in eq.~\eqref{eq:SHcoh}, we cannot
project out these matter fields using Wilson lines. This is why we
choose not to turn on any Wilson lines in the hidden sector.

To summarize, this alternative hidden sector can be used if one wants
to work in the weakly or strongly coupled heterotic string. It does
not rely on five-branes to cancel the anomaly. The hidden $E_8$ gauge
group is broken to $\Spin(12)$, and there are two matter fields
transforming in the $\Rep{12}$ of $\Spin(12)$. These are, of course,
uncharged under the visible $SU(3)\times SU(2)\times U(1)^2$ gauge
group.


\appendix

\section{Pushdown Formulae for Invariant Line Bundles on a
  \texorpdfstring{$\mathbf{dP_9}$}{dP9} Surface}
\label{sec:pushdownformula}

In the course of the computation of the cohomology groups, we need to
know the pushdown of invariant line bundles on a $dP_9$ surface $B$
with elliptic fibration $\beta:B\to\CP^1$ (that is, either $B_1$ or
$B_2$). The invariant degree-$2$ cohomology is $2$-dimensional and
generated by divisor classes $f$ and $t$,
\begin{equation}
  H^2\Big(B,\, \Q\Big)^\ZZZ = \Z f \oplus \Z t
  \,.
\end{equation}
Therefore, all line bundles on $B$ are of the form $\Osheaf_B(n f + m
t)$, $n,m\in\Z$. But the fiber class $f$ is, by definition, just the
preimage of a point on the base $\CP^1$. By the projection
formula, we immediately conclude that
\begin{equation}
  R^k \beta_\ast \Osheaf_B\big( n f + m t) 
  = 
  \Big( R^k \beta_\ast \Osheaf_B\big( m t) \Big) 
  \otimes \OsheafP(n)
  \,,\quad
  n,m\in\Z
  \,.
\end{equation}
Combined with relative duality, eq.~\eqref{eq:relativeduality}, we
find
\begin{equation}
  \beta_\ast \Osheaf_B\big( n f ) = \OsheafP(n)
  \,,\qquad
  R^1 \beta_\ast \Osheaf_B\big( n f ) = 
  K_{B|\CP^1}^\dual \otimes \OsheafP(n)  
  \,.
\end{equation}

Pushing down line bundles that are not vertical is more
complicated. For that, recall that $t$ is the sum of three distinct
sections of the elliptic fibration,
\begin{equation}
  t = \xi + \alpha\xi + (\eta\boxplus\xi)
  \,.
\end{equation}
For convenience, we list the intersection numbers of these sections in
Table~\ref{tab:intersection}.
\begin{table}[htbp]
  \centering
  \begin{tabular}{c|ccc}
    $\bullet$
    &
    $\displaystyle \xi $ & 
    $\displaystyle \alpha \xi $ & 
    $\displaystyle \eta \boxplus \xi $ 
    \\ \hline
    $\displaystyle \xi $ & 
    $-1$ & $1$ & $0$
    \\
    $\displaystyle \alpha \xi $ & 
    $1$ & $-1$ & $1$
    \\
    $\displaystyle \eta \boxplus \xi $ &
    $0$ & $1$ & $-1$
  \end{tabular}
  \caption{Intersection table on the $dP_9$}
  \label{tab:intersection}
\end{table}
Now each section $s$ of the elliptic fibration is, by definition, also
a divisor. Hence, for each line bundle $\Osheaf_B(D)\in \Pic(B)$,
there is a short exact sequence of sheaves
\begin{equation}
  0 
  \longrightarrow
  \Osheaf_B(D-s)
  \longrightarrow
  \Osheaf_B(D)
  \longrightarrow
  \Osheaf_s(D\cdot s)
  \longrightarrow
  0
\end{equation}
coming from the restriction to the (complex) codimension-one variety
$s\subset B$. For example, choosing $s=D=\xi$, we obtain
\begin{equation}
  0 
  \longrightarrow
  \Osheaf_B
  \longrightarrow
  \Osheaf_B(\xi)
  \longrightarrow
  \Osheaf_\xi(-1)
  \longrightarrow
  0  
  \,.
\end{equation}
The associated long exact sequence for the pushdown is then
\begin{equation}
  0 
  \longrightarrow
  \OsheafP
  \longrightarrow
  \beta_\ast \Big( \Osheaf_B(\xi) \Big)
  \longrightarrow
  \OsheafP(-1)
  \stackrel{\delta}{\longrightarrow}
  \OsheafP(-1)
  \longrightarrow
  0  
  \,.  
\end{equation}
The coboundary map $\delta$ has to be an isomorphism for degree
reasons and, therefore,
\begin{equation}
  \beta_\ast \Big( \Osheaf_B(\xi) \Big)  
  = \OsheafP
  \,.
\end{equation}
We can use this method to inductively find the pushdown as we add one
section at a time. For example, for $\Osheaf_B(\xi\oplus \alpha\xi)$
we find a short exact sequence
\begin{equation}
  0 
  \longrightarrow
  \Osheaf_B(\xi)
  \longrightarrow
  \Osheaf_B(\xi+\alpha\xi)
  \longrightarrow
  \Osheaf_{\alpha\xi}
  \longrightarrow
  0  
  \,,
\end{equation}
which pushes down to the short exact sequence 
\begin{equation}
  0 
  \longrightarrow
  \OsheafP
  \longrightarrow
  \beta_\ast\Big( \Osheaf_B(\xi+\alpha\xi) \Big) 
  \longrightarrow
  \OsheafP
  \longrightarrow
  0  
\end{equation}
which determines unambiguously the pushdown to be
\begin{equation}
  \beta_\ast\Big( \Osheaf_B(\xi+\alpha\xi) \Big) 
  = 2 \OsheafP
  \,.
\end{equation}
Continuing this way, we can, in principle, determine the pushdown for
all $\Osheaf_B(nt)$. However, eventually one encounters extension
ambiguities that make it more difficult to find a unique
answer. Happily, one can avoid the extension ambiguities up to
$\Osheaf_B(5t)$ if one just adds sections in the correct order. 
Reading from left to right, the preferred order to add the sections is 
\begin{equation} 
\begin{split}
  t =&\, 
  \xi + \alpha\xi + (\eta \boxplus \xi)
  \\
  2 t =&\, 
  \xi + \alpha\xi + (\eta \boxplus \xi) + 
  \alpha\xi + \xi + (\eta \boxplus \xi)
  \\
  3 t =&\, 
  \xi + \alpha\xi + (\eta \boxplus \xi) + 
  \alpha\xi + \xi + \alpha\xi + 
  \xi + (\eta \boxplus \xi) + (\eta \boxplus \xi)    
  \\
  4 t =&\, 
  \xi + \alpha\xi + (\eta \boxplus \xi) + 
  \alpha\xi + \xi + \alpha\xi + 
  \\ &\,+
  \xi + (\eta \boxplus \xi) + \alpha\xi + 
  \xi + (\eta \boxplus \xi) + (\eta \boxplus \xi)    
  \\
  5t =&\,
  \xi + \alpha\xi + (\eta \boxplus \xi) + 
  \alpha\xi + \xi + \alpha\xi +
  \\ &\,+
  \xi + (\eta \boxplus \xi) + \alpha\xi + 
  (\eta \boxplus \xi) + \alpha\xi + \xi + 
  \xi + (\eta \boxplus \xi) + (\eta \boxplus \xi)  
  \,.
\end{split}
\end{equation}
The corresponding pushdown is then easily determined to be
\begin{equation}
  \begin{split}
    \beta_\ast\Big( \Osheaf_B(t) \Big) &=\, 
    3 \OsheafP
    \\
    \beta_\ast\Big( \Osheaf_B(2t) \Big) &=\, 
    6 \OsheafP
    \\
    \beta_\ast\Big( \Osheaf_B(3t) \Big) &=\, 
    8 \OsheafP \oplus \OsheafP(1)
    \\
    \beta_\ast\Big( \Osheaf_B(4t) \Big) &=\, 
    9 \OsheafP  \oplus 3\OsheafP(1)
    \\
    \beta_\ast\Big( \Osheaf_B(5t) \Big) &=\, 
    9 \OsheafP  \oplus 6\OsheafP(1)
    \,.
  \end{split}
\end{equation}
One can also deal with the extension ambiguities for
$\beta_\ast\Osheaf(nt)$, $n>5$, and arrive at a unique
answer. However, this is not necessary for our purposes.

\section{Direct Image of an Extension}
\label{sec:pushdownext}

The purpose of this Appendix is to show that the pushdown
map 
\begin{multline}
  \label{eq:pushdownExtinv}
  \beta_{2\ast}:~
  \Ext^1\Big( 
    \chi_2 \OsheafBtwo(2t+f)\otimes I_9 ,~
    \OsheafBtwo(2t-3f)
  \Big)^G
  \longrightarrow
  \\
  \longrightarrow
  \Ext^1\bigg( 
    \underbrace{
      \beta_{2\ast}\Big( \chi_2 \OsheafBtwo(2t+f)\otimes I_9 \Big)
    }_{=3 \OsheafP(-1)\oplus 3\OsheafP} ,~
    \underbrace{
      \beta_{2\ast}\Big( \OsheafBtwo(2t-3f) \Big)
    }_{=6 \OsheafP(-3)}
  \bigg)^G
\end{multline}
of the $G=\ZZZ$ invariant part of the Ext groups is the zero map. To
conclude this, we will look at the local to global spectral sequence
for the Ext groups. For the rest of this Appendix, let $\HOM$ and
$\EXT$ be the local Hom and Ext, that is, the sheaf of homomorphisms
and the corresponding derived functor. Note that a quick application
of the long exact sequence associated to the short exact sequence of
sheaves, eq.~\eqref{eq:I9ses}, yields
\begin{equation}
  \EXT^i\Big( I_9, \OsheafBtwo \Big) = 
  \begin{cases}
    \OsheafBtwo\,, & i=0 \\
    \bigoplus_{i=1}^{9} \Osheaf_{p_i}\,, & i=1 \\
    0\,, & \text{otherwise}\,.
  \end{cases}  
\end{equation}
Using this and elementary properties of the local extensions,
see~\cite{Hartshorne}, we find
\begin{equation}
  \EXT^i\Big( 
    \chi_2 \OsheafBtwo(2t+f)\otimes I_9 ,~
    \OsheafBtwo(2t-3f)
  \Big)
  = 
  \begin{cases}
    \chi_2^2 \OsheafBtwo(-4f)\,, & i=0 \\
    \bigoplus_{i=1}^{9} \Osheaf_{p_i}\,, & i=1 \\
    0\,, & \text{otherwise}\,.    
  \end{cases}
\end{equation}
Of course we are only interested in the global $\Ext$ which classifies
extensions of sheaves. The global $\Ext$ groups are determined from
the local Ext sheaves by means of a spectral sequence. In general, for
arbitrary sheaves $\Esheaf$ and $\Fsheaf$ this spectral sequence
starts with
\begin{equation}
  E_2^{p,q} = 
  H^p\Big( \EXT^q\big( \Esheaf, \Fsheaf \big) \Big) 
  ~\Longrightarrow~
  \Ext^{p+q}\big( \Esheaf, \Fsheaf \big)
  \,.
\end{equation}
In our case, the local to global spectral sequence
\begin{multline}
  E_2^{p,q}\big(B_2 \big) = 
  H^p\Big( 
  \EXT^q\Big( 
    \chi_2 \OsheafBtwo(2t+f)\otimes I_9 ,~
    \OsheafBtwo(2t-3f)
  \Big)
  ~\Longrightarrow\\\Longrightarrow~
  \Ext^{p+q}\Big( 
    \chi_2 \OsheafBtwo(2t+f)\otimes I_9 ,~
    \OsheafBtwo(2t-3f)
  \Big)  
\end{multline}
for the sheaves on $B_2$ starts with
\begin{equation}
  E_2^{p,q}\big(B_2 \big) = 
  \vcenter{\xymatrix@=3mm{
      {\scriptstyle q=1}\hspace{1.6mm} &
      \Reg(G)
      \ar[drr]^(0.25){d_3} & 
      0 & 0 \\
      {\scriptstyle q=0}\hspace{1.6mm} &
      0 & 
      \chi_2^2+\chi_1\chi_2^2+\chi_1^2\chi_2^2 &
      \chi_2^2+\chi_1\chi_2^2+2 \chi_1^2\chi_2^2 \\      
      \ar[]+/r 3.4mm/+/u 1.7mm/;[rrr]+/r 15mm/+/r 3.4mm/+/u 1.7mm/
      \ar[]+/r 3.4mm/+/u 1.7mm/;[uu]+/u  2mm/+/r 3.4mm/+/u 1.7mm/
      & 
      {\vbox{\vspace{3.5mm}}\scriptstyle p=0} & 
      {\vbox{\vspace{3.5mm}}\scriptstyle p=1} &  
      {\vbox{\vspace{3.5mm}}\scriptstyle p=2} 
    }}
  \,.
\end{equation}
The $d_3$ differential is nontrivial. One way to determine it is to
compare with the result for the global $\Ext^1$ group, see
eq.~\eqref{eq:ExtWresult}. From that, we can conclude that the image
of $d_3$ has to be $\chi_2^2+\chi_1\chi_2^2+\chi_1^2\chi_2^2$ and,
hence, the third (and final) tableau must be
\begin{equation}
  E_\infty^{p,q}\big(B_2 \big) = 
  \vcenter{\xymatrix@=3mm{
      {\scriptstyle q=1}\hspace{1.6mm} &
      (1+\chi_2)(1+\chi_2+\chi_2^2)
      & 
      0 & 0 \\
      {\scriptstyle q=0}\hspace{1.6mm} &
      0 & 
      \chi_2^2(1+\chi_1+\chi_1^2) &
      \chi_1^2\chi_2^2 \\      
      \ar[]+/r 3.4mm/+/u 1.7mm/;[rrr]+/r 3mm/+/r 3.4mm/+/u 1.7mm/
      \ar[]+/r 3.4mm/+/u 1.7mm/;[uu]+/u  2mm/+/r 3.4mm/+/u 1.7mm/
      & 
      {\vbox{\vspace{3.5mm}}\scriptstyle p=0} & 
      {\vbox{\vspace{3.5mm}}\scriptstyle p=1} &  
      {\vbox{\vspace{3.5mm}}\scriptstyle p=2} 
    }}
  \,.
\end{equation}
On the other hand side, the local Ext of the direct image sheaves is
\begin{equation}
\begin{split}
  \EXT^i\Big( 
    3 \OsheafP(-1)\oplus 3\OsheafP ,~
    6 \OsheafP(-3)
  \Big)
  =  
  \hspace{3cm} \\ \hspace{3cm}
  =
  \begin{cases}
    18 \OsheafP(-2) \oplus 18 \OsheafP(-3) \,, & i=0 \\
    0\,, & \text{otherwise}\,.    
  \end{cases}
\end{split}
\end{equation}
The corresponding local to global spectral sequence then has only one
nonvanishing entry and looks schematically like
\begin{equation}
  E_2^{p,q}\big(\CP^1 \big) = 
  E_\infty^{p,q}\big(\CP^1 \big) = 
  \vcenter{\xymatrix@=3mm{
      {\scriptstyle q=1}\hspace{1.6mm} &
      0 & 0 & 0 \\
      {\scriptstyle q=0}\hspace{1.6mm} &
      0 & 
      \cdots &
      0 \\      
      \ar[]+/r 3.4mm/+/u 1.7mm/;[rrr]+/r 3mm/+/r 3.4mm/+/u 1.7mm/
      \ar[]+/r 3.4mm/+/u 1.7mm/;[uu]+/u  2mm/+/r 3.4mm/+/u 1.7mm/
      & 
      {\vbox{\vspace{3.5mm}}\scriptstyle p=0} & 
      {\vbox{\vspace{3.5mm}}\scriptstyle p=1} &  
      {\vbox{\vspace{3.5mm}}\scriptstyle p=2} 
    }}
  \,.
\end{equation}
But the pushdown image of the $G$-invariant part of the global Ext
groups, eq.~\eqref{eq:pushdownExtinv}, has to be induced from a map of
spectral sequences. But because it has to respect the $(p,q)$ degree,
there is no non-zero map
\begin{equation}
  E_2^{p,q}\big(B_2)^G 
  ~\stackrel{0}{\longrightarrow}~
  E_2^{p,q}\big( \CP^1 \big)^G
  \,.
\end{equation}
Hence, the map eq.~\eqref{eq:pushdownExtinv} is the zero map, as
claimed.

\section{Other Equivariant Actions}
\label{sec:otherequiv}

There is a different way to distribute the characters occurring in our
standard model vector bundle. For that, we change the rank $2$ bundle
$\Vsheaf_1$ to $\Vsheaf_1'$, defined as 
\begin{equation}
  \Vsheaf_1'
  \eqdef
  \OsheafXt(-\tau_1+\tau_2) \oplus 
  \chi_2^2
  \OsheafXt(-\tau_1+\tau_2) 
  \,.
\end{equation}
The rank $4$ vector bundle $\Vsheaf'$ is again defined as a generic
extension of the form
\begin{equation}
  0 
  \longrightarrow
  \Vsheaf_1'
  \longrightarrow
  \Vsheaf'
  \longrightarrow
  \Vsheaf_2
  \longrightarrow
  0
  \,.
\end{equation}
If one were only to look at the underlying holomorphic vector bundles,
then $\Vsheaf$ and $\Vsheaf'$ are the same bundle\footnote{This is not
  quite correct: $\Vsheaf_1'$ and $\Vsheaf_1$ are the same underlying
  bundle, but the extensions $\Vsheaf'$ and $\Vsheaf$ are different,
  because one must choose different extensions if one wants the
  extension to be equivariant. Indeed, stable holomorphic bundles have
  uniquely determined equivariant structures, up to multiplication by
  an overall phase.}. But as equivariant bundles they differ, because
the equivariant $\ZZZ$ action is different. Hence, the quotient
$\Vsheaf/(\ZZZ)$ and $\Vsheaf'/(\ZZZ)$ are different vector bundles on
$X=\Xt/(\ZZZ)$.

This is illustrated by the cohomology groups. For example, consider
the cohomology of $\wedge^2\Vsheaf$ versus $\wedge^2\Vsheaf'$. Of
course, the dimension is the same since they are equal as ordinary
vector bundles,
\begin{equation}
  \dim_C 
  H^1\Big( \Xt,~ \wedge^2\Vsheaf \Big) = 
  14 =
  \dim_C 
  H^1\Big( \Xt,~ \wedge^2\Vsheaf' \Big)  
  \,.
\end{equation}
But the $14$ dimensional $\ZZZ$ representation differs. One can easily
show that
\begin{equation}
    H^1\Big( \Xt, \wedge^2 \Vsheaf \Big)
    =
    2 \oplus 
    2\chi_1 \oplus 
    2\chi_2 \oplus
    2\chi_1^2 \oplus
    2\chi_2^2 \oplus 
    2\chi_1\chi_2^2 \oplus 
    2\chi_1^2\chi_2     
    \,,
\end{equation}
whereas
\begin{equation}
    H^1\Big( \Xt, \wedge^2 \Vsheaf' \Big)
    =
    2              \oplus 
    \chi_1         \oplus 
    \chi_1^2       \oplus 
    2\chi_2        \oplus
    2\chi_1\chi_2  \oplus
    \chi_1^2\chi_2 \oplus 
    2\chi_2^2      \oplus
    \chi_1\chi_2^2 \oplus
    2\chi_1^2\chi_2^2
    \,.
\end{equation}
We can now read off the spectrum if one were to use the different
bundle $\Vsheaf'$ for compactification, and obtain one pair of Higgs,
two $\Rep{3}$, and two $\barRep{3}$ multiplets. Again, doublets and
triplets are split but not in the desired way.

\bibliographystyle{JHEP} \renewcommand{\refname}{Bibliography}
\addcontentsline{toc}{section}{Bibliography} \bibliography{main}

\end{document}